\documentclass{article} 
\usepackage{colm2024_conference}

\usepackage{booktabs}
\usepackage{graphicx}
\usepackage{enumitem}
\usepackage{wrapfig}
\usepackage{algorithm}
\usepackage{algpseudocode}
\usepackage{natbib}
\usepackage{makecell}
\usepackage{booktabs}
\usepackage{array}
\usepackage{amsmath}
\usepackage{amssymb}
\usepackage{amsfonts}
\usepackage{multirow}
\usepackage{verbatim}
\usepackage{caption}
\usepackage{longtable}
\usepackage{supertabular}
\usepackage{CJKutf8}
\usepackage[utf8]{inputenc} 
\usepackage[T1]{fontenc}
\usepackage[french,vietnamese,mongolian,greek,english]{babel}
\usepackage{pifont}

\usepackage{enumitem}
\usepackage{tablefootnote}
\usepackage{xspace}
\usepackage{textcomp}
\usepackage{makecell}
\usepackage{lscape} 
\usepackage{siunitx}
\usepackage{listings}
\usepackage{xcolor}

\usepackage{setspace}   

\usepackage{scrextend}

\usepackage{tgpagella}
\usepackage{latexsym}
\usepackage[T1]{fontenc}
\usepackage[utf8]{inputenc}
\usepackage{microtype}
\usepackage{url}            
\usepackage{nicefrac}       
\usepackage{changepage}
\usepackage{xargs}          
\usepackage{wrapfig,lipsum,booktabs}
\usepackage{longtable}
\usepackage{subcaption}
\usepackage{endnotes}

\usepackage{pgfplots}
\usetikzlibrary{pgfplots.groupplots}
\pgfplotsset{compat=1.3}
\usepackage{tikz}
\usetikzlibrary{patterns}

\usepackage[most]{tcolorbox}

\usepackage[capitalize,noabbrev]{cleveref}
\crefname{section}{Section}{\S\S}
\Crefname{section}{Section}{\S\S}
\crefname{table}{Table}{Tables}
\crefname{figure}{Figure}{Figures}
\crefname{algorithm}{Algorithm}{}
\crefname{equation}{eq.}{}
\crefname{appendix}{Appendix}{}
\crefformat{section}{Section #2#1#3}
\usepackage{multicol}
\usepackage{tcolorbox}
\usepackage{titlesec}
\titleformat*{\section}{\large\bfseries}

\usepackage[utf8]{inputenc} 
\usepackage[T1]{fontenc}    
\usepackage{url}            
\usepackage{booktabs}       
\usepackage{amsfonts}       
\usepackage{nicefrac}       
\usepackage{microtype}      
\usepackage{xcolor}         
\usepackage{mdframed}
\usepackage{tcolorbox}

\usepackage{amsmath}
\usepackage{globalvals}
\usepackage{graphicx}
\usepackage{wrapfig}
\usepackage{diagbox}
\usepackage{wrapfig}
\usepackage{multirow}
\usepackage{tabularx,colortbl}
\usepackage{soul}
\usepackage{float}

\usepackage{hyperref}       

\newcommand{\vpara}[1]{\vspace{0.06in}\noindent\textbf{#1 }}
\newcommand{\vvpara}[1]{\vspace{0.06in}\noindent\textit{#1 }}

\newcommand{\model}{MOON\xspace}

\title
{\model Embedding: Multimodal Representation Learning for E-commerce Search Advertising}

\author{
\vspace{-5mm}
\setstretch{1.2}    
Chenghan Fu\thanks{Equal Contribution. Please refer to Sec.~\ref{sec:contributors} for the complete list of contributors to this work.}, \quad  
Daoze Zhang\footnotemark[1], \quad 
Yukang Lin\footnotemark[1], \quad 
Zhanheng Nie\footnotemark[1], \quad
Xiang Zhang, \quad  Jianyu Liu, \quad Yueran Liu, \quad
Wanxian Guan, \quad Pengjie Wang, \quad Jian Xu, \quad Bo Zheng
\\ 
\texttt{ \footnotesize
\{fuchenghan.fch, zhangdaoze.zdz, linyukang.lyk, niezhanheng.nzh, genshen.zx, liuyu.ljy, tianer.lyr, wanxian.gwx, pengjie.wpj, xiyu.xj\}@taobao.com, bozheng@alibaba-inc.com
} \\
\vspace{2mm}
\textbf{Search Advertising Team, Alimama (Alibaba Group)}
\vspace{-3mm}
}

\begin{document}

\maketitle

\setstretch{1.2}    

\vspace{-7mm}

\begin{abstract}

\textbf{Achievements.}
We introduce \model~\footnote{
MOON is the abbreviation for \underline{\textbf{M}}ultim\underline{\textbf{o}}dal Representation Learning for E-commerce Pr\underline{\textbf{o}}duct U\underline{\textbf{n}}derstanding.
}, our comprehensive set of sustainable iterative practices for multimodal representation learning for e-commerce applications.
\model has already been fully deployed across all stages of Taobao search advertising system, 
including retrieval, relevance, ranking, and so on.
The performance gains are particularly significant on click-through rate (CTR) prediction task, which achieves an overall +20.00\% online CTR improvement. 
Over the past three years, this project has delivered the largest improvement 
on CTR prediction task and undergone five full-scale iterations.
Throughout the exploration and iteration of our \model, 
we have accumulated valuable insights and practical experience that we believe will benefit the research community.

\vpara{Insights.}
The integration of multimodal content into CTR prediction model has been widely recognized as essential.
As early as 2022, we anticipated that the future of CTR prediction would rely on multimodal understanding, ushering in a next-generation paradigm that combines sparse and dense modeling.
At that time, the mainstream idea in the industry followed a simple end-to-end paradigm, jointly training multimodal and CTR models by combining multimodal representations with complex feature interactions.
We also experimented with this method, but after identifying its limitations, we analyzed and concluded that such end-to-end training struggles to reach the full potential.
Therefore, we pioneered a multi-stage, decoupled integration approach between multimodal representations and downstream models, which ultimately led to the significant advancements reported above.

\vpara{Contributions.}
Along this new avenue, our \model contains a three-stage training paradigm of ``Pretraining, Post-training, and Application'', allowing effective integration of multimodal representations with downstream tasks. 

Notably, to bridge the misalignment between the objectives of multimodal representation learning and downstream training, 
we define the \textit{exchange rate} to quantify how effectively improvements in an intermediate metric can translate into downstream gains. 
Through this analysis, we identify the image-based search recall as a critical intermediate metric guiding the optimization of multimodal models.

Over three years and five iterations, \model has evolved along four critical dimensions: data processing, training strategy, model architecture, and downstream application.
The lessons and insights gained through the iterative improvements will also be shared.

As part of our exploration into scaling effects in the e-commerce field, 
we further conduct a systematic study of the scaling laws governing multimodal representation learning for CTR prediction, examining multiple factors such as the number of training tokens, negative samples, and the length of user behavior sequences.

Finally, we develop an efficient and scalable infrastructure that spans the entire life cycle of multimodal representations, from their production to their consumption, addressing various challenges inherent in our multi-stage paradigm, 
including storage management, data I/O, computation efficiency, and real-time perception.

\end{abstract}
\newpage
\tableofcontents  
\newpage
\section{Introduction}\label{sec:intro}

We introduce \model, our comprehensive set of sustainable iterative practices for multimodal representation learning for e-commerce applications.
\model has already been fully deployed across all stages of Taobao search advertising system, 
including retrieval, relevance, ranking, and so on.
The performance gains are particularly significant on click-through rate (CTR) prediction task, which achieves an overall +20.00\% online CTR improvement. 
Over the past three years, this project has delivered the largest improvement 
on CTR prediction task and undergone five consecutive full-scale iterative upgrades.
A natural question arises: Why were we able to establish a sustainable path and achieve such significant gains?
Throughout the exploration and iteration of our \model, we have accumulated a wealth of experience and insights. We believe these findings are worth sharing with the community, and this technical report serves that purpose~\footnote{
As the first installment of a series, this report focuses on our practice of multimodal representation learning for CTR prediction in the coarse and fine ranking scenarios. Future reports to be released will continue to elaborate on its broader applications in more downstream tasks like retrieval and relevance modeling.
}.

With the rapid development of large language models (LLMs) and multimodal large language models (MLLMs), the scenarios such as recommendation systems, web search, and online advertising have increasingly sought to integrate large models in recent years, with multimodal content-based CTR (Click-Through Rate) prediction serving as a prominent example. 
Incorporating multimodality into CTR prediction is essential because users’ interaction behaviors are basically driven by visually engaging content, such as product images and videos, rather than by textual information alone. 
Motivated by this observation, as early as 2022, based on our assessment of industry trends, we anticipated and firmly believed that CTR prediction would inevitably converge with multimodal understanding, forming the next-generation CTR estimation paradigm that integrates both sparse and dense models. 
Yet, 
in the practical implementation and subsequent five iterations, 
we encountered several key challenges that shaped our understanding, with details as follows.

In 2022, the mainstream idea in the industry 
adopted a straightforward end-to-end paradigm,
i.e., combining relatively simple multimodal representations with complex downstream feature interactions, and jointly training of the multimodal model and the CTR model,
and we also experimented with this approach.
However, after encountering some of its drawbacks, through systematic analysis and reflection, we realized that the end-to-end idea for multimodal-content-based CTR prediction would likely struggle to reach its performance ceiling. 
This limitation primarily stems from the significant heterogeneity between multimodal content and CTR models.
(1) 
In terms of convergence characteristics, 
the sparse ID embeddings used in downstream CTR models exhibit sharp distributional variability. 
Also, these models often converge within only one epoch~\citep{zhang2022towards} and place high demands on the temporal freshness of the training data.
By contrast, multimodal dense representations are more stable in distribution, usually require many more training steps, and do not demand time-sensitive data.
(2) 
With respect to the module architecture, 
the performance improvement of sparse CTR models primarily depends on feature interactions, whereas dense multimodal models emphasize more on architectural design. 
Therefore, joint optimization not only struggles to balance these two requirements effectively, but also greatly increases training complexity.
(3) 
As for the modeling objects, CTR models focus on capturing user interest through click behaviors, while multimodal models concentrate on understanding textual and visual content, leading to a large gap between the two modeling spaces.
Therefore, the limited effectiveness of such joint training methods can be mainly attributed to the fundamental heterogeneity between multimodal models and CTR ones,
which makes seamless integration inherently difficult.


Based on this analysis, we argue that achieving superior performance in multimodal-content-based CTR prediction requires a multi-staged, decoupled integration of multimodal representations and downstream models, instead of naive joint training. 
As a result, we immediately shifted and paved a new avenue for a multi-stage training paradigm, focusing on optimizing representation quality, which allows us to achieve the accomplishments described above. 
Considering that existing public efforts in industry remain highly insufficient in the direction of multi-stage training, we believe that sharing our experiences will be valuable to the community, mainly including in the following aspects.

\begin{itemize}
    \item 
From the perspective of supervision signals, under a multi-stage training paradigm, 
the optimization objective for the multimodal model is not identical to the ultimate downstream target. 
Therefore, it is extremely crucial to identify an appropriate intermediate metric that is strongly and monotonically correlated with downstream performance. 
However, there still remains a lack of public, well-substantiated conclusions on how to select such an intermediate metric as the objective for training multimodal representations.

    \item 
From the viewpoint of sustainable iteration, considering that a decoupled architecture allows the multimodal model to be optimized independently, its representation is expected to benefit from scaling laws as data volume and model scale increase. 
Yet, detailed guidance of how to iteratively upgrade the multimodal components such that their scaling gains can reliably translate into better CTR performance in downstream applications, are still very scarce.

    \item
From the implementation efficiency perspective, a multi-stage pipeline introduces longer processing chains that can substantially increase 
computational load, 
storage cost, and engineering complexity, 
posing severe challenges for actual online deployment. 
Practical accounts on achieving low latency in large-scale real-world applications are rarely shared; and efficient engineering solutions suitable for production environments are basically underreported.

\end{itemize}

\begin{figure}[h]
  \centering
  \includegraphics[width=\linewidth / 100 * 100]{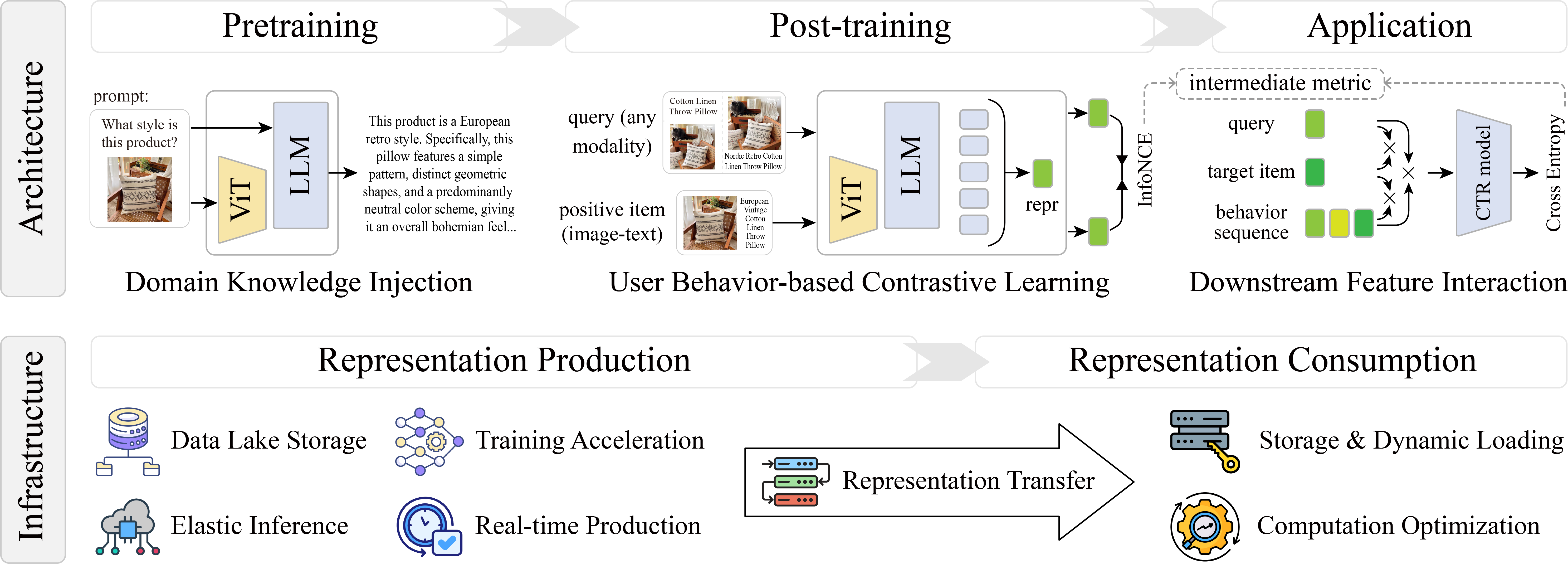}

  \caption{The overview of the architecture and infrastructure of our \model's latest iteration. 
  } 
  \label{fig:framework}
\end{figure}

Therefore, this report introduces \model, a comprehensive \textit{set of sustainable iterative practices} for \textbf{m}ulti\textbf{o}dal pr\textbf{o}duct u\textbf{n}derstanding for CTR prediction (Fig.~\ref{fig:framework}),
which includes: 
(1) the three-stage architecture and training paradigm of ``Pretraining, Post-training, and Application''; 
(2) the explored insights derived from our five successive iterations, covering advances in data quality, training strategy, model architecture, and scale expansion; and 
(3) the engineering solution that ensures efficiency in training and deployment, including the representation production and consumption.
The comparison between the joint end-to-end training idea and our \model is given in Tab.~\ref{tab:comparison_exsiting}.

In this report, we mainly present the following significant findings:

\begin{itemize}
    \item 
We propose and validate the three-stage architecture of ``Pretraining, Post-training, and Application'', identifying the \textit{image-based search recall} as the critical intermediate metric that bridges multimodal representation learning with downstream objectives, which is highly essential for maintaining optimization consistency in decoupled training.

    \item 
Through five iterative versions of our \model, we continuously improve our work along four key dimensions: training strategy, model architecture, data processing, and downstream application. We also systematically explore the scaling laws of various expansion factors that govern multimodal representation learning for CTR prediction.

    \item
We develop an efficient, scalable, and comprehensive infrastructure solution that covers the entire life cycle of multimodal representations, from their production to their consumption, addressing a wide range of challenges inherent in the multi-stage paradigm, including storage management, data I/O, computational efficiency, and real-time perception.

    \item 
Experimentally, our \model not only achieves strong results on the intermediate metric but also delivers an overall +20.00\% improvement on CTR prediction in online A/B test. 
Furthermore, the visualizations of retrieval tasks and image–text alignment provide additional evidence of our model’s powerful capability for multimodal product content understanding.

\end{itemize}

In the remainder of this report, in Sec.~\ref{sec:method}, we introduce the technical details of the latest fifth version of our model architecture and training method.
Sec.~\ref{sec:infrastructure} details our infrastructure solution for training and deployment efficiency. 
In Sec.~\ref{sec:eval}, we first present the intermediate performance metric, image-based search retrieval recall, which links representation learning to downstream application objectives, 
and then report the experiment performance on the CTR prediction task in both offline and online scenarios. 
At the end of Sec.~\ref{sec:eval}, we 
share our empirical exploration of scaling laws in many aspects for multimodal CTR modeling. 
Finally, Sec.~\ref{sec:history} summarizes the trajectory of our work, including lessons learned from iterating the training data, multimodal representations, and the CTR prediction model.

\begin{table}[ht]
    \caption{Comparison between the joint and our decoupled training for multimodal CTR prediction.}
    \label{tab:comparison_exsiting}
    \centering
    \footnotesize
    \renewcommand{\arraystretch}{1.2} 
    \setlength\tabcolsep{13pt}
    
    \begin{tabular}{ccc}
    \toprule
     & \textbf{End-to-end Training} & \textbf{Multi-stage Pipeline} \\
    \midrule
    \textbf{Existing Methods} & \begin{tabular}[c]{@{}c@{}} EM3~\citep{deng2024end} \\ 
    MEDA~\citep{fan2024multi}  \\
    Diff-MSIN~\citep{cui2025diffusion}
    \end{tabular} 
    & our \model \\
    
    \cmidrule(lr){1-3}
    
    \textbf{Architecture} & \begin{tabular}[c]{@{}c@{}} Open-source MLLM \\ with joint training \end{tabular}  
    & \begin{tabular}[c]{@{}c@{}} ``Pretraining, Post-training  \\ and Application'' \end{tabular}    \\

    \cmidrule(lr){1-3}

    \textbf{Improvement Focus} & Feature Interaction & Representation Capabilities \\

    \cmidrule(lr){1-3}
    
    \textbf{Representation Model} 
    & \begin{tabular}[c]{@{}c@{}} 
        Relatively Small \\
        (EM3: 0.1B parameters \\ trained on 411B tokens) 
    \end{tabular} 
    
    & \begin{tabular}[c]{@{}c@{}} Large Scale \\ (4B  trained on 2400B tokens) \end{tabular} \\

    \cmidrule(lr){1-3}
    
    \textbf{CTR Prediction Model} 
    & \begin{tabular}[c]{@{}c@{}} Heavy, end-to-end encoding  \\ with feature interaction \end{tabular} 
    & \begin{tabular}[c]{@{}c@{}} Light-weight Attention \end{tabular} \\

    \cmidrule(lr){1-3}
    
    \textbf{Intermediate Metric} & Not Required & Search Recall \\
    \bottomrule
    \end{tabular}

\end{table}

\section{Method }\label{sec:method}

\subsection{Overview}

In Sec.~\ref{sec:method}, we present the technical details of the 
latest fifth iteration
of this work. 
As discussed in Sec.~\ref{sec:intro}, considering the huge differences between multimodal models and CTR models in terms of convergence characteristics, architectural design, and modeling objects, we adopt a multi-stage training paradigm rather than a coupled approach for building multimodal content understanding models for CTR prediction. 
Specifically, we decompose the problem into two primary learning objectives: (1) generative-model-based multimodal representation learning, and (2) downstream user visual preference modeling with the content-based user behavior extractor (CUBE). 
The ideas and methods behind accomplishing these two objectives will be described in Sec.~\ref{sec:2.1.1} and Sec.~\ref{sec:2.1.2}, respectively. 
Then, Sec.~\ref{sec:pretraining}, Sec.~\ref{sec:posttrain}, and Sec.~\ref{sec:application} will provide more detailed explanations of the data construction, data processing, model architectures, and training strategies at each stage of the training process.

\subsubsection{Generative MLLM-based Multimodal Representation
Learning } 
\label{sec:2.1.1}

In fact, before the work of our \model, 
we had previously attempted to introduce multimodal features into CTR prediction models, but initially observed no significant benefits.
As mentioned in Sec.~\ref{sec:intro}, our diagnosis is that our prior methods underestimated the importance of the capabilities of multimodal representations themselves, 
concentrating on fusing sparse and dense features within the CTR predictor. 
Subsequently, we revisited the problem of multimodal content-based CTR prediction and, based on the analysis in Sec.~\ref{sec:intro}, shifted our focus toward optimizing high-quality multimodal representations.
The specific details are described in the following paragraphs and elaborated in Sec.~\ref{sec:pretraining} and Sec.~\ref{sec:posttrain}.

\begin{figure}[h]
  \centering
  \includegraphics[width=\linewidth / 100 * 70]{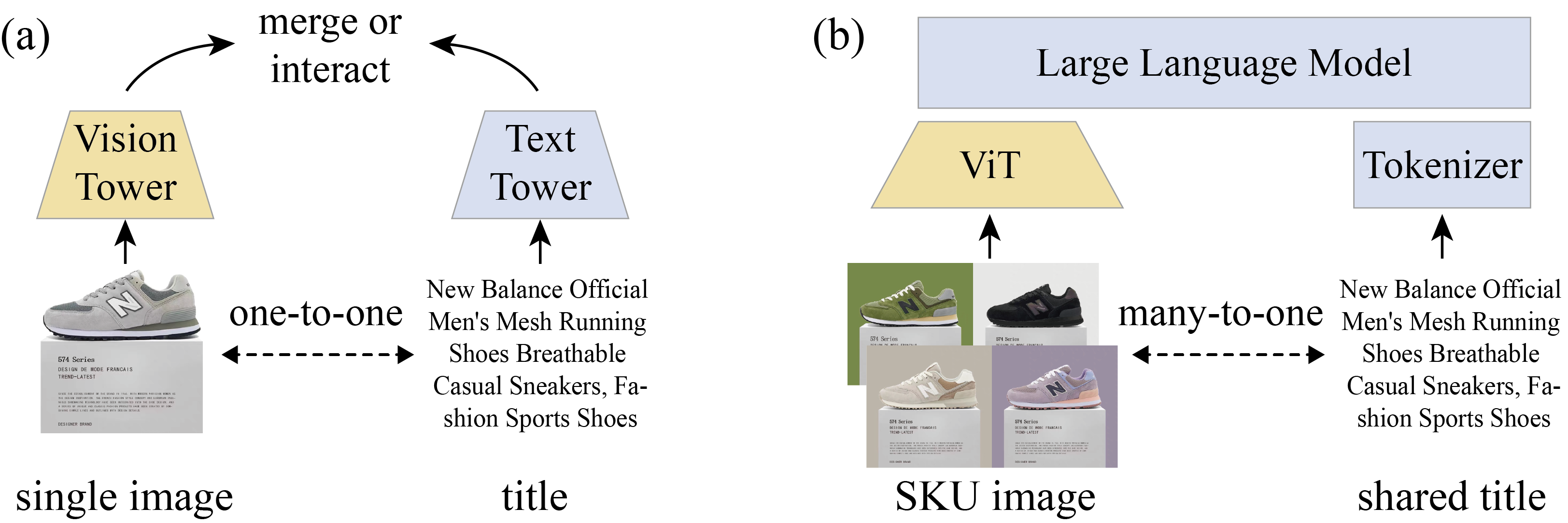}

  \caption{Comparison between the dual-flow and MLLM architectures. 
  \small \setstretch{1.2}
  (a) The dual-flow paradigm is inherently limited to encode one-to-one image-text pairs and cannot directly capture many-to-one relationships. 
  (b) The MLLM-based idea is naturally suited to model the richer visual content from multiple SKU images.
  } 
  \label{fig:data_1}
\end{figure}

For multimodal representation learning of products, many existing works~\citep{dai2024uniembedding,dong2022m5product,jin2023learning,liu2023multimodal,zhan2021product1m} focus on adopting the dual-encoder (also called dual-flow) architecture that employs a visual encoder and a text encoder to separately process multimodal content for each product. 
While such approaches are intuitive and propel progress in the field, as illustrated in Fig.~\ref{fig:data_1}, they are mainly based on a one-to-one modeling paradigm, which fails to directly model the many-to-one relationship between multiple images of the same product (such as stock-keeping unit (SKU) images) and their shared title. 
By contrast, in general multimodal fields, some works~\citep{lin2024mm,zhang2024gme} have explored multimodal large language models (MLLMs)~\citep{liu2024improved,wang2024qwen2,zhang2025sharper} for multimodal representation learning, which offer greater flexibility in modeling richer information from multiple images or texts.

Therefore, inspired by the employment of MLLMs for general multimodal representation learning, 
the most recent iteration of our \model has innovatively adopted generative MLLMs for product content understanding. 
To effectively leverage generative models for product-level multimodal representation learning, three core challenges need to be addressed:

\begin{itemize}
    \item 
\textbf{Domain Knowledge in E-commerce.} Open-source MLLMs are typically trained for broad, general scenarios and may provide limited support for e-commerce entities, attributes, images, and linguistic patterns. Therefore, equipping the foundation model with domain knowledge is greatly essential for high-quality product representation learning.

    \item
\textbf{Transferring Autoregression to Representation.} A generative MLLM is usually trained for next token prediction, while our objective is to yield product representations,
which shows a clear gap in task formulation. 
So, 
how to transfer the generative MLLMs to yield strong representations that can be applied to downstream applications, is a significant challenge.

    \item
\textbf{Learning Discriminative Representations.} In the e-commerce field, product catalogs span numerous categories and styles, where inter-item differences can be very subtle. 
The multimodal model must capture and understanding nuanced similarities and differences across various products, enabling strong discrimination that supports downstream tasks. 

\end{itemize}

To address these challenges, we progressively train different capabilities of the model across the pretraining and post-training stages, culminating in a multimodal model that yields strong, robust, and discriminative representations for e-commerce products.
Specifically, our training of multimodal representation for product content comprises:

\begin{itemize}
    \item 
For pretraining, we employ an inner-developed MLLM, named TBStars-VL, as our generative foundation model. 
The TBStars-VL model is pretrained with rich e-commerce domain knowledge, forming the foundation for reliable product representation learning. (Details in Sec.~\ref{sec:pretraining}.)

    \item
In the post-training stage, to bridge the gap between the autoregressive generation and our goal of representation learning, 
we first finetune the model using diverse real-world user behavior data.
Through a contrastive learning task, this stage teaches our model to produce a set of product representations of multiple sizes, allowing the adaptability across heterogeneous downstream scenarios with different latency and memory budgets. (Details in Sec.~\ref{sec:posttrain_1}.)

    \item 
To strengthen the finer understanding for products, we collect and process higher-quality purchase data and conduct a further training stage. 
Here we specially enlarge the pool and difficulty of negative samples during the contrastive training, 
such that the model can capture subtle inter-item differences and yields robust, discriminative representations.
(Details in Sec.~\ref{sec:posttrain_2}.)
    
\end{itemize}

\subsubsection{User Visual Preference Modeling for Downstream Tasks}
\label{sec:2.1.2}

After the pretraining and post-training, the product representations are applied as feature embeddings in CTR prediction and more downstream tasks. 
As discussed in Sec.~\ref{sec:intro} and Sec.~\ref{sec:2.1.1}, we believe that the improvement in multimodal-content-based CTR prediction depends primarily on 
learning high-quality multimodal representations, which play a more crucial role than the complexity of feature interactions within the CTR predictor.
Therefore, our downstream CTR model treats the product representations in a simple and efficient manner within the content-based user behavior extractor (CUBE), which contains two main steps: 
(1) First, to fully utilize the core advantages of multimodal representations,
we compute the similarity between each target item and the user behavior sequence in the latent space, where we adopt cosine similarity, which is consistent with the InfoNCE~\citep{oord2019representation} loss used during multimodal training. 
(2) Then, we perform element-wise multiplication between the sequence of similarity scores and the original sequence of ID embeddings to achieve complementary fusion with the content semantic information.
More details about the multimodal features are provided in Sec.~\ref{sec:application}.

In summary, our latest \model introduces a three-stage paradigm of ``Pretraining, Post-training, and Application'', 
which sequentially fulfills two learning objectives—generative-model-based multimodal representation learning and downstream user visual preference modeling—thereby showing stronger performance on multimodal-content-based CTR prediction and more downstream tasks.

\subsection{Pretraining}\label{sec:pretraining}

For pretraining, we employ an internally developed MLLM, TBStars-VL, as our generative foundation model. 
Unlike open-source general MLLMs, TBStars-VL is pretrained not only on large-scale general corpora, but also on substantial e-commerce data to strengthen its domain understanding. 
These additional training data involve multiple e-commerce-oriented question answering and grounding tasks, including product description generation, visual grounding of product images, attribute extraction, and product question answering. 
This targeted pretraining endows the generative MLLM with adequate knowledge in e-commerce, thereby providing a strong foundation for subsequent finetuning and allowing reliable learning of product-centric multimodal representations for downstream business tasks.

\subsection{Post-training} \label{sec:posttrain}

\subsubsection{From Autoregression to Representation Learning} \label{sec:posttrain_1}

Following the pretraining stage, our model has acquired a solid understanding of e-commerce-specific knowledge. 
However, its generative nature remains confined to the next token prediction paradigm, which is far from our objective of representation learning. 
To bridge this gap, in the first stage of post-training, we transform the autoregressive output into the ability to yield meaningful representations. 
Specifically, we utilize the user-behavior-driven multimodal contrastive learning, leveraging real-world multimodal retrieval scenarios as supervision. 
Through full-parameter finetuning under this setting, the model is encouraged to align visual and textual modalities while simultaneously learning to extract product representations at multiple granularities (dimension sizes), thereby establishing the basic capability for product representation learning.

\begin{figure}[t]
  \centering
  \includegraphics[width=\linewidth / 100 * 100]{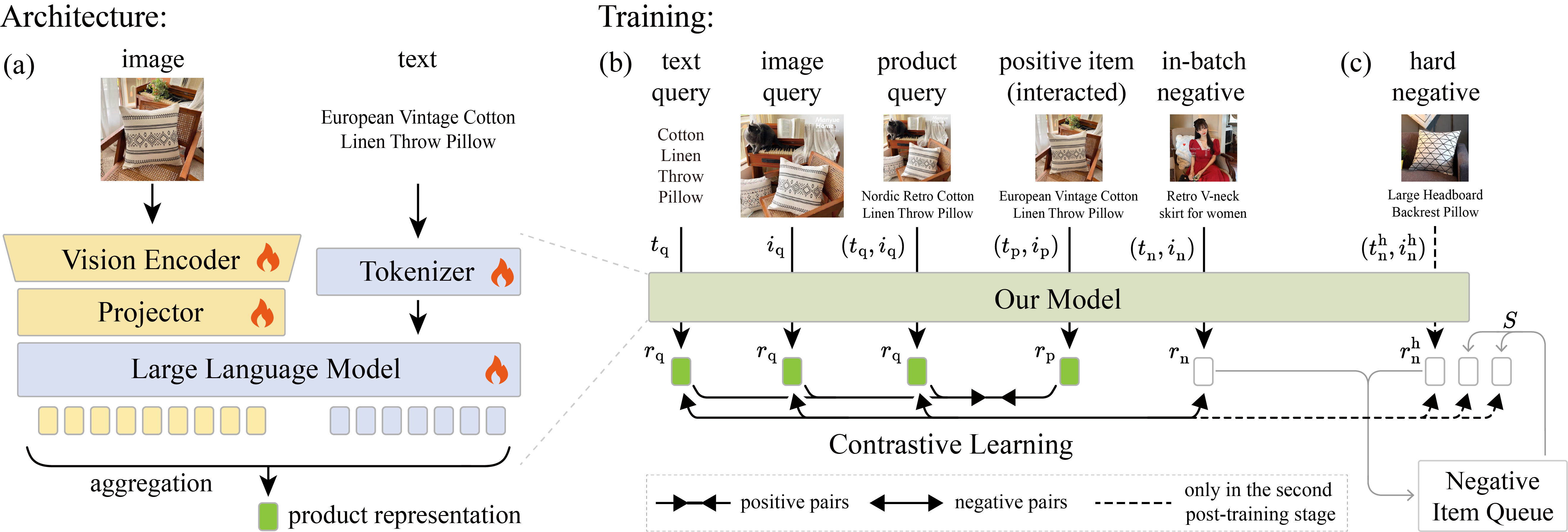}

  \caption{The architecture and training of our \model model. 
  \small \setstretch{1.2}
  (a) 
  Beyond the dual-encoder paradigm, we employ the generative-MLLM-based method for product representation learning. 
  (b) 
  In post-training, we leverage real-world user behaviors as supervision to effectively capture latent correlations between related product items. 
  (c) 
  Moreover, during the second post-training stage, we not only construct hard negative samples, but also use the Spatial-Temporal Negative Sampling, 
  to learn more robust and discriminative representations.
  } 
  \label{fig:model_arch}
\end{figure}

\vpara{Model Architecture.}
As illustrated in Fig.~\ref{fig:model_arch}(a), to convert the autoregressive generation into representation output, the input mirrors the common practice of MLLMs: a vision encoder consumes product images, and an LLM processes the associated text. 
On the output side, 
the hidden states from the LLM’s last layer are aggregated by mean pooling to obtain a sentence-level embedding, which serves as the final representation of the input products or queries. 
Under this architecture,
our model naturally accepts arbitrary input modalities, including text-only, image-only, or image–text, exposing a unified embedding interface. 
This unified formulation enables support for multiple downstream retrieval scenarios. 

Since our objective is multimodal representation learning rather than causal generation, the unidirectional attention mask commonly used in LLMs is unnecessary. 
Therefore, we replace it with bidirectional attention, allowing each token to attend to the full context, 
which fully enhances the model's ability to understand the visual and textual contents of products.

Furthermore, to learn representations that are both powerful and versatile across a wide range of downstream business applications, 
we aim to provide representations that can accommodate diverse resource and performance constraints. 
Instead of producing a single fixed-size vector, our \model model yields a set of representations at multiple dimensionalities: 128, 256, 512, 1024, and 3072. 
To achieve this, we incorporate Matryoshka Representation Learning (MRL)~\citep{kusupati2022matryoshka} during training, which introduces supervision signals across multiple latent spaces. 
This technique ensures that lower-dimensional embeddings preserve as much semantic information as possible, thereby alleviating the information loss typically occurring in feature compression. 
In downstream applications, the scenarios with enough computational and storage resources can exploit the richer and more expressive higher-dimensional representations; while resource-constrained or latency-sensitive applications can rely on the representations with smaller size that still retain strong semantic fidelity.

\vpara{Contrastive Finetuning.} 
For the fine-tuning task, as shown in Fig.~\ref{fig:model_arch}(b), we use a contrastive learning objective, where the supervision signal is based on the real-world user behaviors, including clicks, favorites, add-to-cart events, and purchases.
We mine valuable user behavior logs to construct positive training pairs, where an item actually interacted with by the user is treated as a positive sample for a given query. 
Formally, we use $t_\text{q}$ or $i_\text{q}$ to denote a text query or image query, 
$(t_\text{p}, i_\text{p})$ to denote the image and text content of a positive sample, and $(t_\text{n}, i_\text{n})$ to denote the image and text of a negative sample.
During the contrastive training, our model encodes the query 
$q \in \{ t_\text{q}, i_\text{q}, (t_\text{q}, i_\text{q}) \}$, 
its corresponding positive item $(t_\text{p}, i_\text{p})$, and the negative item $(t_\text{n}, i_\text{n})$ sampled from the mini-batch, to learn their representations $r_\text{q}$, $r_\text{p}$ and $r_\text{n}$, respectively.
Finally, we adopt the InfoNCE loss to maximize the mutual information between the query $r_\text{q}$ and its positive item $r_\text{p}$:
\begin{equation}
\mathcal{L}_1 = 
-\log 
\frac{\exp\left(
r_\text{q}^\intercal \cdot r^{}_\text{p} / \tau
\right)}
{
\exp\left(r_\text{q}^\intercal \cdot r^{}_\text{p} / \tau \right) + 
\sum_{r^{'}_\text{n} \in S } 
\exp\left(
r_\text{q}^\intercal \cdot r^{'}_\text{n} / \tau
\right)
},
\end{equation}
where $\tau$ denotes the temperature hyperparameter to adjust scale, 
$S$ denotes the set of negative samples,
and $\mathcal{L}_1$ denotes the InfoNCE loss between the query and the positive item.

\vpara{Data Construction and Ratio.} 
In the fine-tuning described above, all positive pairs for contrastive learning are derived from real-world user interactions. 
Specifically, for each search event on the Taobao platform, we form a binary pair consisting of the query and an interacted positive item.
Positive interactions include clicks, favorites, add-to-cart events, and purchases.
Here we consider three retrieval scenarios, i.e., text-based search, image-based search, and product-based search, so the query can be text-only, image-only, or an image–text input.
After collecting raw behavior logs, we filter out the noisy data with extremely short or meaningless text. 
This curation yields approximately 2.6 billion (query, positive item) pairs in total. 
The distribution of these samples across various cross-modal retrieval scenarios is given in Tab.~\ref{tab:data_dist}.

\begin{table}[h]
    \caption{Training data ratio for the first stage of finetuning. 
    \small
    Here, the ``richer-text'' contains the product's textual attributes, category, descriptions and so on. 
    }
    \label{tab:data_dist}
    \centering
    \footnotesize
    \setlength\tabcolsep{4.5pt}
    \setstretch{1.2}
    
    \begin{tabular}{cccccccccc}
    \toprule
    \textbf{Query}          & text        & text  & text          & text                & image & image         & image               & \begin{tabular}[c]{@{}c@{}} title \\ + image \end{tabular}   & \begin{tabular}[c]{@{}c@{}} richer-text \\ + image \end{tabular} \\
    \midrule
    \textbf{Item}           & richer-text & image & \begin{tabular}[c]{@{}c@{}} title \\ + image \end{tabular} 
    & \begin{tabular}[c]{@{}c@{}} richer-text \\ + image \end{tabular} 
    & image & \begin{tabular}[c]{@{}c@{}} title \\ + image \end{tabular} 
    & \begin{tabular}[c]{@{}c@{}} richer-text \\ + image \end{tabular}
    & \begin{tabular}[c]{@{}c@{}} title \\ + image \end{tabular} & \begin{tabular}[c]{@{}c@{}} richer-text \\ + image \end{tabular} \\
    \midrule
    \textbf{\#Samples}         & 0.2B         & 0.2B   & 0.2B           & 0.5B                 & 0.2B   & 0.5B           & 0.2B                & 0.5B & 0.1B          \\
    \bottomrule
    \end{tabular}
\end{table}

\subsubsection{High-Quality Representation Finetuing} \label{sec:posttrain_2}

After contrastive training with diverse positive user behaviors, the next token prediction  of our model has been successfully transformed into representation learning. 
However, the model's ability to extract product semantics remains relatively shallow. 
To further enhance the modeling and understanding of fine-grained product semantics, 
we conduct an additional round of full-parameter finetuning using high-quality real-world purchase data. 
This is because, compared with common interactions like clicks or favorites, purchase behavior inherently implies stronger and more reliable semantic correlations between products, 
thereby serving as a richer supervision signal for capturing subtle inter-item distinctions.
In addition to the mentioned techniques in the earlier post-training stage, 
this stage introduces stricter data-cleaning pipelines and more advanced contrastive training strategies, ensuring more robust, generalizable and discriminative representations. 
Detailed explanation is given as follows.

\vpara{Data Construction and Filtering.} 
To further improve the model’s capacity to learn fine-grained product semantics, this stage uses only purchase behaviors to construct positive pairs for contrastive learning. 
Although other interactions (like clicks and favorites) are more abundant, they imply weaker positive correlations between products. 
In contrast, confirmed purchases provide a more faithful and discriminative signal of inter-item correlations, and thus are a better supervision for representation learning. 
Specifically, we mine real-world purchase logs to form (query, purchased item) pairs, following the same schema as before. 
This initial corpus contains several billion pairs. 
Then, to ensure high data quality, we apply two additional procedures: Similarity-based Deduplication and NER-based Filtering.

\begin{itemize}
    \item 
\textbf{Similarity-based Deduplication.} The raw data still needs to be deduplicated and processed, which mainly includes the following three steps:

\vvpara{(1) Deduplicate the purchased SKU item. }
In e-commerce, despite the large number of purchase transactions, the 
\textit{head effect} means that a small set of popular items accounts for a disproportionate share of traffic and purchases. 
As a result, many raw data samples center on those best-selling products, 
while numerous unpopular items appear only sparsely, 
which is detrimental to uniform improvements in representation quality across the catalog.
Therefore, we need to deduplicate the raw data. 
For multiple data samples that share the same purchased SKU item (with distinct queries), 
we compute the query–item similarity using the ensemble model of our previous iterative models,
and retain only the sample with the lowest similarity score, 
because it is hardest for the model to distinguish its relevance to the purchased SKU item, encouraging the model to learn subtle semantic distinctions and enhancing the generalization ability. 

\vvpara{(2) Merge the SKU item. }
Although we have performed deduplication of purchased items, the uniqueness is guaranteed only at the stock-keeping unit (SKU) level. That is, we only ensure that the SKU ID of each purchased item is unique. 
It is still possible for two pairs to contain the same product as purchased item (just not the same SKU ID, for example, they might be different colors of the same item). 
This may result in pseudo-negative samples during contrastive learning, where a negative sample within the batch is the same product (with different color) as the  positive sample. 
To solve this issue, we further merge the multiple SKU items to one product,
which also helps to obtain richer multi-image information for each product.

\vvpara{(3) Align the distribution to that of CTR samples. }
Through the process above, we obtain a set of (query, purchased item) pairs, in which each purchased item is unique, thereby mitigating the long-tail dominance of popular items and pseudo-negative problem. 
However, the category distribution of purchased items in these pairs still does not match that of the CTR samples\footnote{Here, CTR samples refer to the set of items exposed to all users on the platform over a certain time period, totaling approximately 3 billion samples per day. The category distribution of CTR samples provides an good reflection of the input data distribution that multimodal representations encounter in downstream CTR prediction.}, 
which may hinder the effective application of our representations for downstream CTR prediction. 
To align the data distribution with that of the CTR samples, we first compute the proportion of CTR samples across categories and then adjust the category distribution of the pair set by either supplementing underrepresented categories with new samples or applying random sampling.

    \item
\textbf{NER-based Filtering.} To remove the uninformative samples in the raw data, 
we apply an entity-centric filtering strategy. 
Specifically, we run the e-commerce-oriented Named-Entity Recognition (NER)~\citep{munnangi2024brief} on the textual query and the item’s textual content to extract entity tokens, which captures the product attributes such as color, style, brand, category, and so on. 
Low-quality or meaningless samples typically contain very few entities, so we keep only those samples whose extracted entity count exceeds 2, discarding overly coarse or incomplete data.

\end{itemize}

After this preprocessing, the dataset size is reduced from several billion to approximately 600 million, with marked improvements in data quality, distribution balance, and semantic diversity.

\vpara{Negative-Enhanced Finetuning.} 
Similar to the previous stage (Sec.~\ref{sec:posttrain_1}), this stage also leverages the high-quality data to conduct contrastive learning. 
However, the e-commerce domain involves a vast number of product categories, often numbering in the hundreds of thousands.
Under such conditions, using only basic in-batch negatives is highly likely to sample easy negative items, which diminishes both the efficiency and the effectiveness of contrastive training. 
To address this issue, we aim to not only increase the difficulty of distinguishing between positives and negatives, but also simultaneously expand the diversity and coverage of sampled negatives. 
To this end, as shown in Fig.~\ref{fig:model_arch}(c), we adopt two strategies: Hard Negative Sampling and Spatial–Temporal Negative Sampling~\citep{yan2025mim}, 
enabling the model to learn more generalized, robust, and discriminative product representations.

\begin{figure}[t]
  \centering
  \includegraphics[width=\linewidth / 100 * 90]{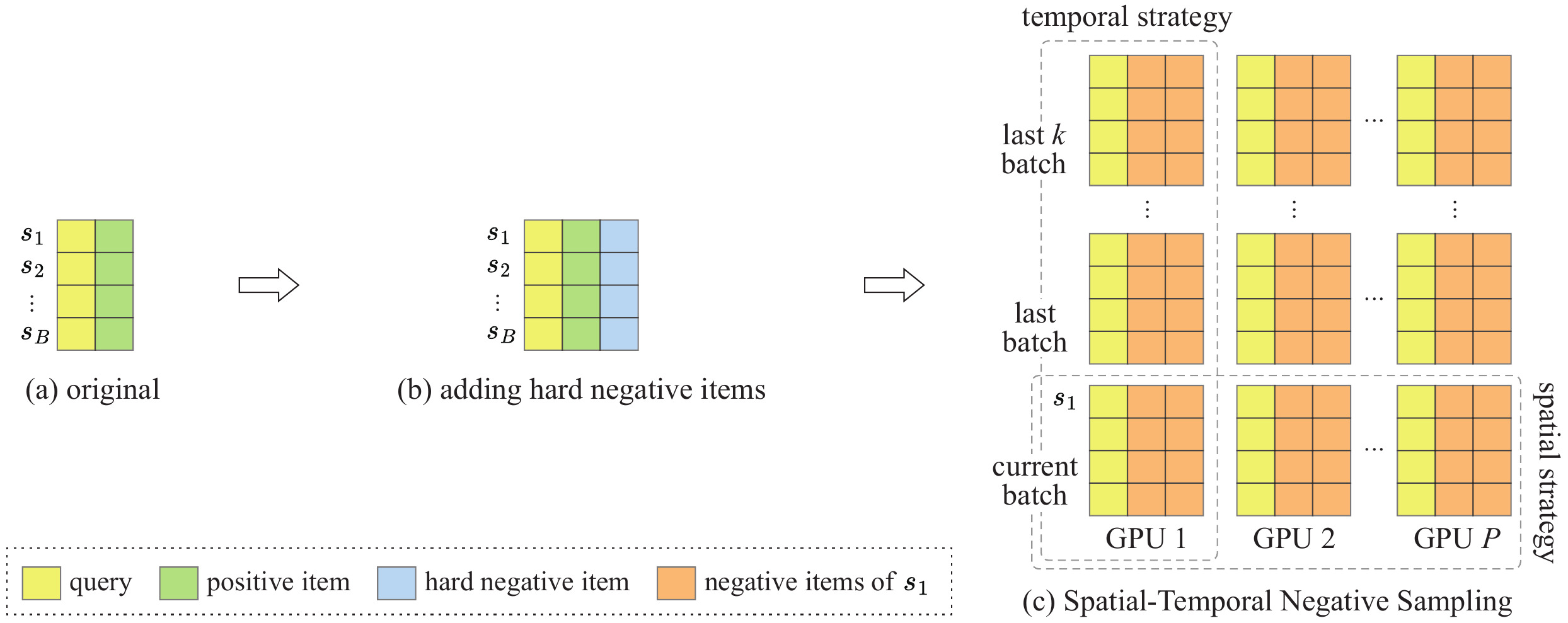}
    
  \caption{Illustration of the Spatial-Temporal Negative Sampling. 
  \small 
  (a) For trivial in-batch sampling, negative items are sampled from other items in the same batch.
  (b) We prepare a similar item belonging to the same category as the query as a hard negative item.
  (c) Taking the sample $s_1$ as an example, we greatly expand the negative pool from both spatial and temporal dimensions.
  }
  
  \label{fig:spatial_temporal_neg}
\end{figure}

\begin{itemize}
    \item 
    \textbf{Hard Negative Sampling.} 
To encourage the model to learn more discriminative representations during contrastive learning, we construct hard negative samples for each query. 
Specifically, for each query, 
we sample a product item that belongs to the same leaf category as the query (but is not the same product) 
as the hard negative sample. 
This transforms each training sample from a (query, positive item) pair into a triplet 
$ ( q, (t_\text{p}, i_\text{p}), (t_\text{n}^\text{h}, i_\text{n}^\text{h})  ) $, 
where $q \in \{t_\text{q}, i_\text{q}, (t_\text{q}, i_\text{q})\}$, and $(t_\text{n}^\text{h}, i_\text{n}^\text{h})$ denotes the sampled hard negative item.

    \item
    \textbf{Spatial-Temporal Negative Sampling.}
Moreover, to further enlarge the pool of negative samples, we adopt two strategies (Fig.~\ref{fig:spatial_temporal_neg}): 
(1) Temporal strategy: 
We store the negative samples of the past $k$ training steps using a item queue. 
Then, the negative samples are collected not only from the current batch but also from the past $k$ batches, increasing the number of negatives to $2B(k+1)-1$, where $B$ is the batch size. 
(2) Spatial strategy: In distributed training, the items from all $P$ GPUs (including all the training nodes) across $k$ recent batches are added as negatives, resulting in $2BP(k+1)-1$ total negatives\footnote{
In our experiment, we set $B$=128, $P$=64, and $k$=10.
}. 
Combined, these strategies provide nearly 640$\times$  more negatives compared to the naive in-batch sampling, significantly improving the model’s ability to distinguish subtle differences between semantically similar products in practice.

\end{itemize}

Finally, we utilize the circle loss~\citep{sun2020circle} to maximize the mutual information between the query $r_\text{q}$ and its positive item $r_\text{p}$:
\begin{equation}
\mathcal{L}_2 = 
\log \left(1 + \exp\left(-\left( r_\text{q}^\intercal \cdot r_\text{p} \right) / \tau \right)\right) 
+ \sum_{r'_\text{n} \in \{r^\text{h}_\text{n}\} \cup \tilde{S}} 
\log \left(1 + \exp\left(\left( r_\text{q}^\intercal \cdot r'_\text{n} \right) / \tau \right)\right),
\end{equation}
where $\tau$ denotes the temperature hyperparameter to adjust scale, $r^\text{h}_\text{n}$ denotes the representation of the hard negative $(t^\text{h}_\text{n}, i^\text{h}_\text{n})$, and $\tilde{S}$ denotes the set of additional negatives obtained via Spatial-Temporal Negative Sampling. $\mathcal{L}_2$ represents the circle loss between the query and the user-purchased positive item.

Here our employed circle loss has several advantages over the InfoNCE loss. Unlike InfoNCE, which requires the computation of the softmax over all negatives, circle loss improves the robustness of contrastive learning by focusing on the relative similarity between the positive pair and the hard negative samples, which results in a more discriminative loss function. 
Moreover, circle loss has a more direct focus on improving the angular margin between positive and negative representations, leading to better separation and improved generalization performance in downstream tasks such as CTR prediction.

\subsection{Application} 
\label{sec:application}

\begin{figure}[t]
  \centering
  \includegraphics[width=\linewidth / 100 * 100]{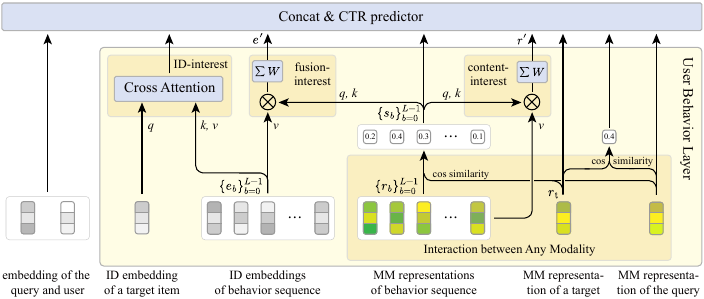}

  \caption{The architecture of content-based user behavior extractor (CUBE) for downstream CTR task. 
  The behavior sequence includes the item behavior sequence and the query behavior sequence.
  } 
  \label{fig:ctr}
\end{figure}

The multimodal representations learned above are fed as features into the downstream models like CTR predictor for business applications. 
It is worth noting that, since our model can yield representations from various modalities, including text, image, and text-image, our representations are suitable for direct interaction across all modalities. 
For example, the query text can interact with the item behavior sequence, and the target item can also interact with the item behavior sequence.
Specifically, as illustrated in Fig.~\ref{fig:ctr}, to fully exploit our multimodal representations in our CUBE, 
we first compute the similarity between the target item $r_\text{t}$ and each item $r_b$ in the user behavior sequence within the learned representation space,
where we use cosine similarity, which is consistent with the InfoNCE objective during multimodal training. 
By this way, we obtain the similarity sequence $\{s_b\}_{b=0}^{L-1}$, that is, the sequence of similarity scores that quantifies the semantic affinity between the target item and each historically interacted item.
\begin{equation}
s_b = 
\frac{r_\text{t} \cdot r_{b}}
{\|r_\text{t}\| \, \|r_{b}\|},
\quad b={0,1,2, ..., L-1}    
\end{equation}
where $L$ denotes the length of the user behavior sequence.

Next, we fuse multimodal semantics with traditional ID embeddings
by multiplying each ID embedding $e_b$ in the behavior sequence with its corresponding similarity score $s_b$, which can be viewed as a similarity-weighted product.
Then, we perform a linear transformation on the obtained sequence and sum it to get the final feature.
\begin{equation}
e' = \sum_{b=0}^{L-1} \text{softmax}(W * s_b) * e_b 
\label{formula:ctr}
\end{equation}
where $e'$ denotes the fused feature of multimodal semantics and ID features. 
Also, similar content-similarity-based weighting operations are performed on the multimodal representation sequence to produce weighted multimodal feature $r'$.
Finally, as shown in Fig.~\ref{fig:ctr}, the aforementioned features 
are fed into the subsequent CTR prediction model.

The method above conveys to the CTR prediction model the content-based similarities between target items and those in the user behavior sequences, measured by cosine similarity of multimodal representations.
This signal
enriches the feature space of the CTR model and contributes to improved prediction accuracy. 
To further enhance the learnability of the cosine similarity value, we perform a linear transformation on the similarity value ($W$ in formula~\ref{formula:ctr}) to obtain the mapped similarities. 
We then use softmax to convert them into a probability distribution that is applied to the ID embeddings $\{e_b\}_{b=0}^{L-1}$ to obtain the weighted sum of the ID embeddings.

\begin{figure}[h]
  \centering
  \includegraphics[width=\linewidth / 100 * 65]{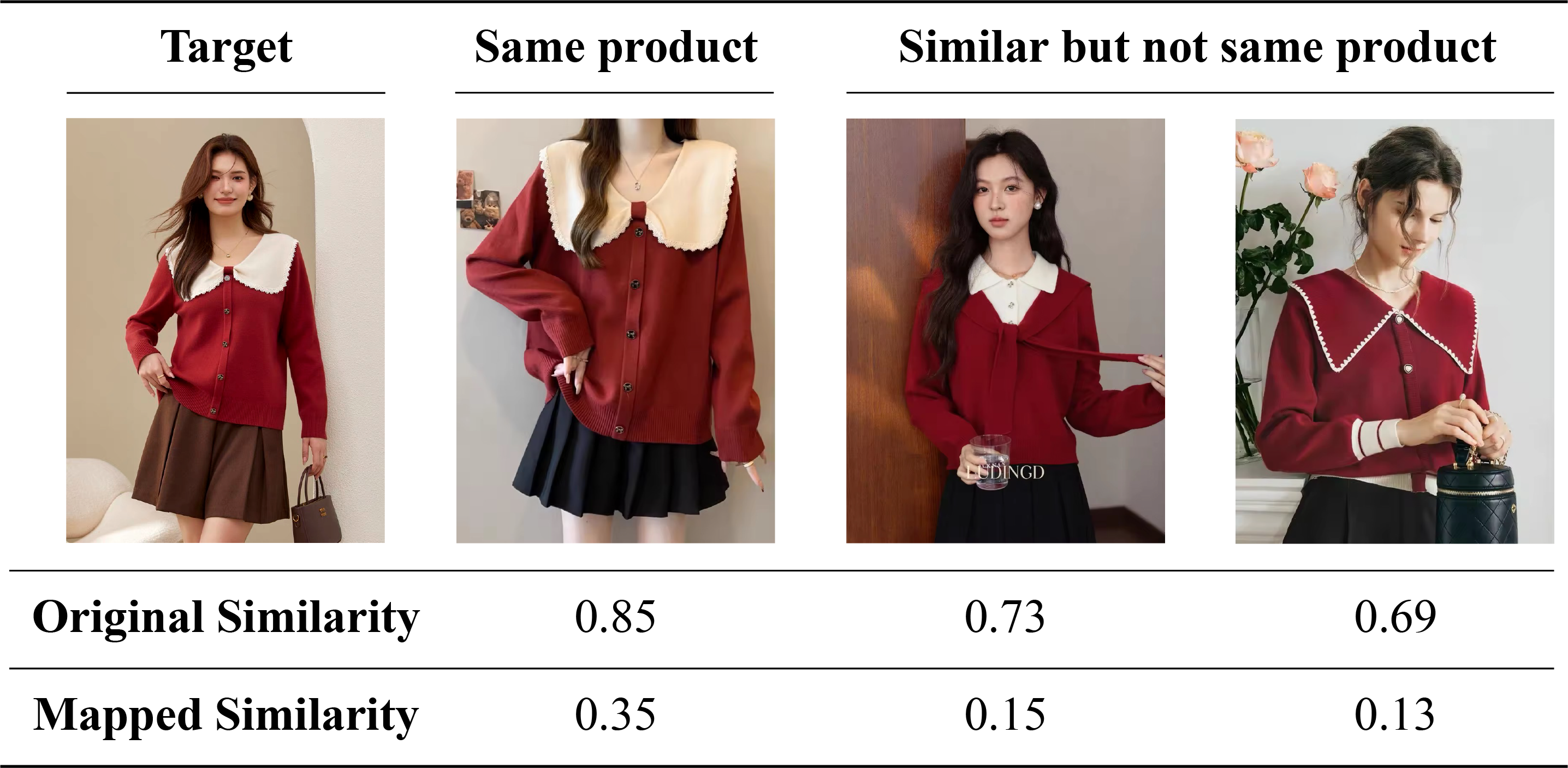}

  \caption{The mapped cosine similarity scores after the linear transformation. 
  \small
  Compared with the original similarities of 0.85 and 0.73/0.69, the mapped similarity of 0.35 shows a much larger difference from 0.15/0.13.
  } 
  \label{fig:sim_enhance}
\end{figure}

The motivation of this linear transformation is that the values of the mapped similarities might be stronger to represent similar but not identical products.
For example, as shown in Fig.~\ref{fig:sim_enhance}, the target and one product (the second column) in the behavior sequence are the same product, 
while the other items (the last two columns) are similar but not identical products. 
Yet, the original cosine similarities, although distinct, remain compressed within a narrow range and fail to highlight the true semantic contrast. 
Consequently, feeding unprocessed cosine scores directly into the CTR predictor is suboptimal.

To address this issue, our linear transformation amplifies the discriminative ability of cosine similarity and produces outputs that more clearly reflect the degrees of content-based similarity. 
This sharper distribution emphasizes highly similar items, facilitates the recognition of similar products, and enables the CTR model to more effectively suppress less similar negative products.
As a result, the transformed similarities convey product-level matches and mismatches with greater clarity.

\section{Infrastructure}
\label{sec:infrastructure}
\begin{figure}[h]

  \centering
  \includegraphics[width=\linewidth ]{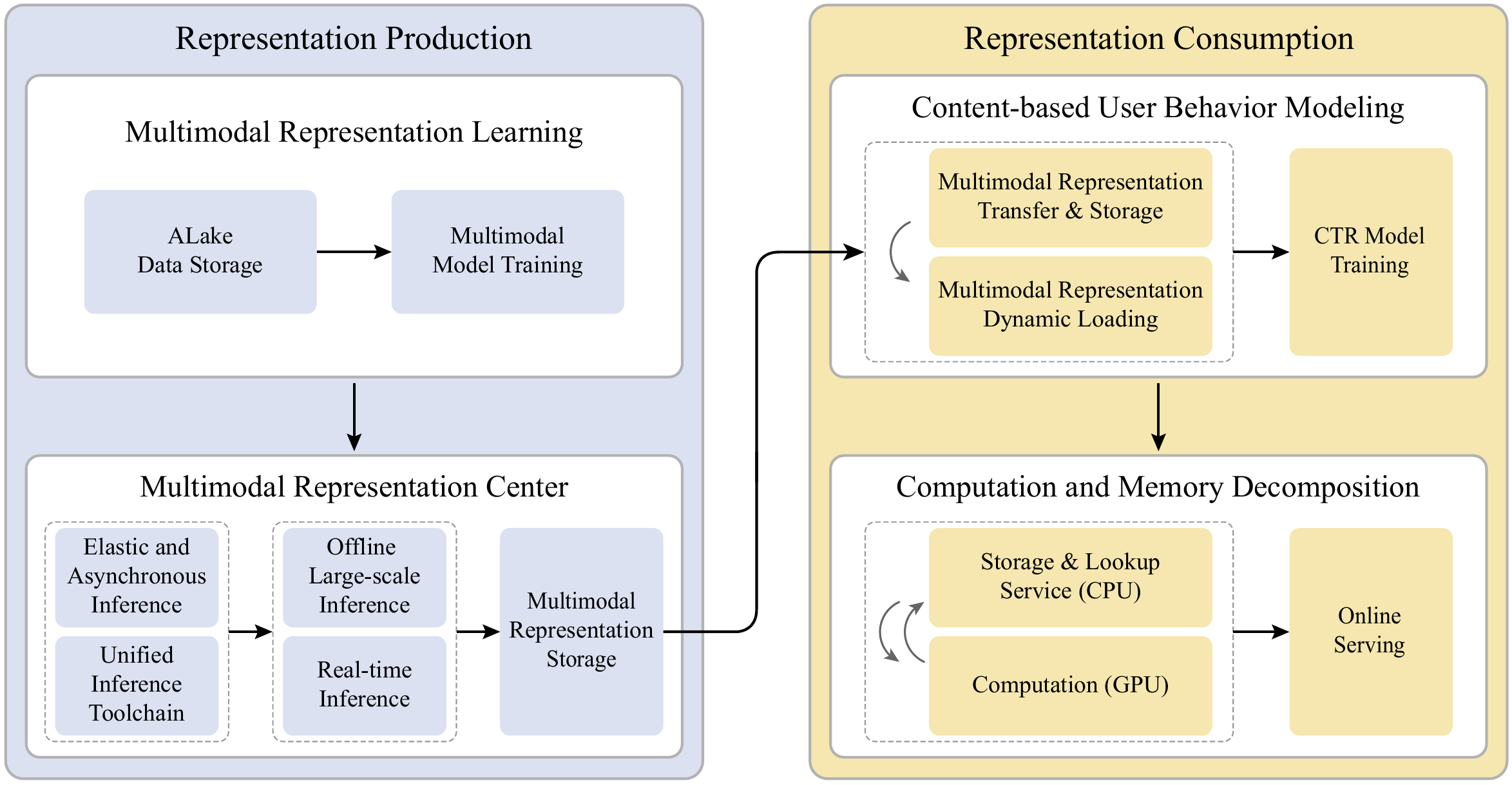}
  \caption{The schematic diagram of the infrastructure engineering optimization.} 
  \label{fig:infra_arc}
\end{figure}

The success of \model, our proposed
multimodal-content-based CTR prediction method,  
hinges not only on algorithmic design, 
but also on a robust, scalable, and efficient infrastructure 
that allows our high-quality multimodal representations to be effectively utilized in downstream industrial applications.
While the previous Sec.~\ref{sec:method} introduces the three-stage model architecture and training paradigm of ``Pretraining, Post-training, and Application'' to decouple the representation learning from downstream applications, 
such a longer pipeline inherently involves many more challenges in storage management, data I/O, computation efficiency, and real-time perception across stages. 
To ensure high efficiency throughout the entire pipeline, 
we establish a engineering solution that spans the entire life-cycle of multimodal representations, from their production in multimodal learning and inference 
to their consumption in downstream training and online serving. 
As demonstrated in Fig.~\ref{fig:infra_arc}, this infrastructure is specifically designed to support sustainable \model model iteration and efficient deployment of multimodal representations, which empasses three key dimensions:

\begin{itemize}
\item 
\textbf{Representation Production.}
To accelerate the training of our multimodal model with billions of parameters
built upon billions of training data, 
we optimize various aspects including data storage, data communication, GPU memory, and computation operators. 
In addition, we design a representation center to support offline large-scale inference and storage of multimodal representations, as well as online real-time inference, ultimately providing efficient query services for downstream tasks.


\item \textbf{Representation Consumption.} 
To ensure efficient utilization of representations in CTR models, we implement optimizations in aspects like representation storage, retrieval, and computation. 

\item \textbf{Real-Time Perception.}
Given strict latency and resource constraints, we devote much effort to ensuring that 
the updated representations from newly uploaded products are propagated to online model inference with near real-time responsiveness.

\end{itemize}

Collectively, these components have laid a solid foundation for the efficient implementation of multimodal representations in CTR model.
In Sec.~\ref{subsec:rep_production}, we first introduce the technical details of the representation production pipeline. Then, the representation consumption pipeline is explained in Sec.~\ref{subsec:rep_consumption}. In addition to the basic two pipelines, we present the approach to support instant recognition of newly added products in Sec.~\ref{subsec:rep_realtime}, preserving the predictive accuracy of the CTR model.

\subsection{Representation Production Pipeline}
\label{subsec:rep_production}

As our model evolved through successive iterations, 
both the parameter size of the multimodal model and the scale of training data also steadily increased (details in Sec.~\ref{sec:history}). 
Consequently, faced with the ever-growing model parameters, computational complexity, data scale, and diverse downstream task requirements, our infrastructure development must continuously evolve to adapt to these changes.
To achieve sustainable iteration of multimodal representations, we design a dedicated representation production pipeline that contains two main stages: (1) efficient model training through system-level optimizations, and (2) 
the construction of an integrated representation center that unifies inference pipelines, large-scale storage, and low-latency querying. 
This framework ensures the efficient, real-time, and low-cost production of multi-modal representations for hundreds of billions of commodities, forming the basis for downstream applications.


\subsubsection{Optimization of Multimodal Representation Learning}
\label{subsubsec:training_optimization}
Multimodal pretraining with billions of parameters on a dataset of billions records introduces significant challenges in data storage, data loading, and computation performance. We tackle these bottlenecks through a combination of storage modernization and deep system-level optimizations.

\vpara{High-Performance Data Storage.}
During training, frequent random access to large volumes of image-text pairs can easily saturate the storage bandwidth. 
Our initial setup relied on a cloud-based Object Storage Service (OSS), which exhibited limitations in query rate (i.e., QPS) and sustained throughput under concurrent workloads, leading to GPU computing power idling and decreased training efficiency. 
To tackle this issue, we design a low-latency, high-throughput stream-batch integrated data lake storage system named ALake, which achieves a significant increase in training I/O speed. 
This system achieves low-latency metadata operations, strong consistency, and scalable read bandwidth by leveraging hierarchical caching and parallelized data paths. 
Empirical evaluation on an A100 GPU shows that this migration reduces end-to-end training time by up to 60\%, primarily by eliminating I/O bottlenecks during data prefetching and shuffling.

\vpara{End-to-End Performance Acceleration.} During the iteration of our multimodal model, the model parameter scale and training data have increased by tens of times, simply scaling up the training resources by the same factor cannot achieve the expected training speed.
To balance training resources and training efficiency, 
we have developed a complete analysis and optimization pipeline through practical experience, including analysis tools and optimization strategies.
First, we improve training efficiency by through standard optimization techniques such as precision quantization, gradient accumulation, and communication strategy optimization.
Then, leveraging analysis tools,
we pinpointed the structures with low-computing power in the model and subsequently performed manual fusion of the underperforming operators. 
The optimization via operator fusion cut down memory accesses per operator from 7 to 3, 
achieving a 2.8$\times$ speedup on the fused unit and a 9\% improvement in end-to-end model throughput.

In summary, the above optimizations ensure that even when the model size and data scale are expanded dozens of times, the total training time can remain almost unchanged without the need to proportionally increase training resources, 
thus supporting the long-term sustainable iteration of the large multimodal model in an industrial-scale setting.

\subsubsection{Computing-in-Memory Multimodal Representation Center }
\label{subsubsec:representation_center}

After the model training, to efficiently deploy the model into downstream tasks, it is necessary to produce multi-modal representations at the hundred billion level. 
To this end, we build a \textit{representation center} for storing billions of multimodal text-image data, allowing offline large-scale batch inference, and online real-time inference for newly added or updated products. Leveraging this platform, we can produce billion-scale representations within 4 days and store them for seamless access by any downstream task.


\vpara{Multimodal Representation Storage.}
For timeliness reasons, rather than directly inferring the multimodal representations during downstream training or inference, we pre-infer representations of all products. 
Then we maintain a centralized table to store multimodal representations, where the key is the product ID and the value is the product multimodal representation. 
In this way, downstream tasks can quickly obtain the representations they need from the complete table of product multimodal representations without any extra GPU resources, inference time or storage overhead.

\vpara{Unified Inference Toolchain.}
To lower the barrier of cross-team cooperation, we design a standardized, component-based pipeline that abstracts the low-level complexities such as data formatting, data preprocessing, and output serialization. 
By this way, given only image URLs and associated text fields, users can simply invoke a unified interface to obtain the normalized multimodal representations. 
This toolchain replaces error-prone manual workflows, such as separate image downloading, format conversion, and script execution, with an automated, version-controlled pipeline. 
It enables accurate delivery of multimodal representations to downstream applications, ensuring consistency across diverse tasks.

\vpara{Elastic and Asynchronous Inference.}
To achieve offline large-scale batch inference, we develop a scalable inference system to manage GPU clusters with auto-scaling capabilities. 
Workloads are scheduled asynchronously, allowing dynamic allocation of computing resources according to demand. 
This elastic mode not only improves system stability under fluctuating loads but also enables efficient employment of spot instances, greatly reducing the operation costs. 
The design effectively decouples inference job submissions from resource availability, ensuring robustness in long-running batch processing tasks.

\vpara{Online Real-time Inference.} 
When a new product is added to industrial applications or an existing product is updated, this module will infer the multimodal representation of these updated products in real-time. 
Using the elastic and asynchronous inference framework ensures that new representations are produced quickly, greatly shortening the time-to-value of new representations and adapting to various throughput requirements while maintaining stable service.

In summary, the representation production pipeline establishes an efficient, scalable, and maintainable system for multimodal learning and inference, laying the foundation for sustainable iteration of high-quality  product multimodal representations.

\subsection{Representation Consumption Pipeline}
\label{subsec:rep_consumption}

Since the production of multimodal representations is decoupled from CTR model training, their reliable, efficient, and timely integration into both offline training and online serving within downstream CTR prediction is critical to the overall effectiveness of the system.
In Sec.~\ref{subsec:rep_consumption}, we detail our engineering solutions for the representation consumption pipeline, which includes two major components: 
training optimization for downstream model and inference acceleration for online serving.

\subsubsection{Training Optimization for Downstream Models}
\label{subsubsec:training_optimization}
While separating representation learning from CTR modeling offers greater flexibility for richer content and has a higher performance ceiling (Sec.~\ref{sec:intro}), 
it also creates new challenges in feature consumption, system compatibility, and resource efficiency. 
Specifically, the upstream representations must be reliably and efficiently delivered downstream to the CTR trainers, and dynamically updated in response to changes in product content, and be consumed under strict computational and memory constraints. 
To address these requirements, we conducted a systematic optimization of the downstream training framework across four key dimensions: (1) multimodal representation transfer, (2) multimodal representation dynamic loading, (3) I/O optimization, and (4) memory optimization. These advancements jointly support the efficient utilization of multimodal representations during offline CTR model training.

\vpara{Multimodal Representation Transfer and Storage.}
Generally speaking,
dense multimodal representations are typically generated using PyTorch-based~\citep{paszke2019pytorch} vision-language models, while large-scale CTR training mainly relies on TensorFlow~\citep{abadi2016tensorflow} frameworks such as XDL~\citep{jiang2019xdl}, 
which are optimized for sparse feature interaction and distributed computation. 
However, the lack of native interoperability between PyTorch and XDL prevents in-memory sharing of dense representations, making it infeasible to stream dense features directly into the training pipeline. 
To enable seamless integration without introducing tight coupling or runtime dependencies, we design a decoupled data mediation mechanism. 
Multimodal representations produced by the PyTorch inference system are persisted in structured tables,
the XDL training framework then reads these tables and converts the representations into serialized model files compatible with sparse model training inputs
This conversion strategy avoids in-time database queries during training, which are prone to bandwidth throttling and increased latency under high concurrency. 
The stability and reproducibility of the multimodal representation inputs have been significantly improved through optimal bandwidth utilization and support for loading multimodal representations with specific versions.

\vpara{Multimodal Representation Dynamic Loading.}
The absence of dynamic multimodal representation updating in the training pipeline can lead to the use of stale features when product content changes, particularly during mid-training, ultimately undermining CTR model accuracy and convergence.
To preserve temporal consistency and freshness of multimodal representations during training, we implement a dynamic loading mechanism within the XDL framework that allows automatic replacement of embedding files at configurable intervals, triggered by wall-clock time or training step count. 
With this mechanism, both full and incremental versions can be loaded seamlessly without interrupting the training process. 
Crucially, our system can maintain strict correspondence between embedding versions and model checkpoints, ensuring that upon task resumption after failure or pause, the exact same feature state is restored. 
This capability not only 
enhances the robustness of the training system but also reduces the need for manual intervention.

\vpara{I/O Optimization. } 
The further profiling of the training pipeline reveals that some CPU-bound operations in the data loading stage severely limit GPU utilization. 
Specifically, legacy implementations perform hash lookups on the CPU during feature fetching, 
which is a computationally expensive operation that introduces significant delay and stalls GPU execution. 
This I/O bottleneck becomes more pronounced when integrating high-cardinality categorical features with dense multimodal representations, leading to underutilized compute resources and prolonged training cycles. 
To solve this issue, we restructured the data ingestion kernel to perform hashing internally within the I/O thread, effectively offloading the computation from the host processor and overlapping it with data transfer. 
Evaluated on the downstream tasks of our \model, this optimization reduces CPU-GPU synchronization overhead and increases end-to-end training throughput by 58\%, 
achieving obviously improved resource efficiency.

\vpara{Memory Optimization.}
The integration of high-dimensional multimodal features significantly increases per-GPU memory consumption, particularly in forward activation storage and intermediate tensor retention. 
Under the HPC synchronous 
~\citep{amodei2016deep} training mode of XDL, maintaining a stable global batch size is essential for convergence stability. 
However, without adjusting per-device batch size to maintain consistency in training settings, the increasing model scale forces expansion of the GPU scale from tens to over 128 nodes, creating unsustainable resource demands that hinder concurrent model development. 
To address this scalability bottleneck, 
we conduct a fine-grained analysis of memory allocation patterns and apply two complementary strategies. 
First, we replace memory-intensive operations such as \texttt{tf.where} with functionally equivalent but lower-footprint alternatives that reduce intermediate tensor creation. 
Second, we employ low-precision int8 multimodal representations to reduc GPU memory usage.
These optimizations enable a 90\% increase in the maximum local batch size (per GPU), allowing effective training under constrained hardware budgets. 
While recomputation incurs a moderate 20\% reduction in per-step speed, the net gain in memory efficiency frees up substantial GPU resources, enabling parallel execution of multiple training jobs and accelerating overall velocity.

In addition to the above optimizations, we are also adopting a more advanced and efficient unified sparse-dense training framework~\citep{zong2025recis} to provide a stronger and more modern infrastructure foundation for the integration of our future multimodal and CTR models.

\subsubsection{Inference Acceleration for Online Serving}

\label{subsubsec:inference_acceleration}
The interaction of dense representations within attention-based modules introduces substantial latency overhead, while the sheer volume of representations exceeds the storage capacity of individual serving instances. 
To ensure low-latency prediction and support scalable deployment, 
we propose a separate strategy where the model parameters (including the representation table and dense layer parameters) and the computation of inference are divided into different machines, i.e., the parameter machine and the computation machine. In this way, the computation engine only needs to account for the memory cost during the feed-forward pass and is not restricted by the large representation table, enabling efficient online inference for CTR models.

\vpara{Storage-Oriented Optimization.}
The total footprint of 10 billion representations of 128 dimensions in FP32 often exceeds 4 terabytes, far surpassing the memory capacity of a single online serving instance. 
Co-locating all representations with the model server leads to severe memory pressure, single-node bottlenecks, and degraded service reliability, particularly during model warm-up or traffic spikes. Moreover, storing full-precision representations at scale imposes prohibitive costs on memory bandwidth and limits the number of concurrent experiments supported. 
To address this issue, we first perform representation accuracy quantification, and then reduce the representation index by removing long-tail items, and finally performed index sharding storage. 
We extend the system with a hierarchical storage architecture (GPU memory, CPU memory, and solid-state drive (SSD)) based on the existing design. 
On the computation nodes, GPU memory serves as the primary storage for representations using the Cuckoo Hash~\citep{pagh2004cuckoo} algorithm, significantly enhancing parallel query throughput, minimizing remote embedding accesses and associated network transmission, and handling approximately 40\% of all queries. 
On the storage nodes, CPU memory serves as the secondary storage while SSD functions as the tertiary storage, jointly support scalable growth in storage capacity.


\vpara{Computation-Oriented Optimization.}
The attention mechanism in CTR models involves intensive vector similarity computations between user and item representations, which become performance-critical when operating on billion-scale multimodal representations. To address this challenge, we performed fine-grained optimization on the computation graph, including: 
(1) rewriting high-cost kernels for hardware-aware execution, 
(2) allowing multi-stream concurrency to overlap computation with communication, and 
(3) implementing dynamic load balancing across heterogeneous compute nodes (e.g., H20 and L20). 
These optimizations achieve a fourfold increase in peak GFLOPS and delivers a 25× improvement in end-to-end inference performance.

\subsection{Real-Time Perception Pipeline}
\label{subsec:rep_realtime}
In the production environments,
delayed propagation of updated representations into training or serving systems leads to feature-label mismatch and calibration undermining, reducing the effectiveness of content-aware personalization. 
To maintain high-quality alignment between user behavior signals and current item semantics, it is essential to minimize the end-to-end latency from content change to representation availability.
This requires the capability that contains not only the speed of generating new multimodal representations but also their real-time integration into downstream pipelines. 
We address this challenge through a coordinated effort across production and consumption stages, focusing on elastic generation infrastructure and incremental indexing update mechanisms.

\subsubsection{Timeliness in Representation Production}
The traditional batch-oriented paradigm for multimodal representation generation often leads to daily update cycles, where new or updated items must wait until the next scheduled inference job to have their representations become available. This delay degrades the overall estimation performance of the entire CTR system, especially during peak listing periods such as promotional campaigns. 
To overcome this limitation, we build an online real-time inference framework (Sec.~\ref{subsubsec:representation_center})  that allows just-in-time execution of representation generation tasks. 

\subsubsection{Timeliness in Representation Consumption}
Even when representations are generated promptly, their impact on model performance remains limited if downstream systems cannot perceive and utilize them in a timely manner. 
In earlier iterations of \model, the propagation delay from representations update to online serving spanned up to one day (T+1), mainly due to offline data dumps and asynchronous index rebuilding processes. 
During this window, CTR models operate on old or missing representations, leading to suboptimal ranking decisions, particularly for time-sensitive items. 
To close this gap, we redesign the consumption path around a fully indexed, incrementally updated, near-real-time embedding serving architecture. 
All generated representations are immediately registered into a distributed embedding index with versioned semantics, enabling immediate lookup upon availability. 
This index is tightly integrated with the online ranking pipeline, allowing the CTR model to access updated representations within seconds of their creation. 
Therefore, the end-to-end perception latency from data update to model visibility is reduced from T+1 to within seconds, ensuring that trained models reflect the most up-to-date content.
This advancement enhances both the modeling of user visual preferences and the feedback loop between representation fidelity and downstream CTR performance.

\section{Evaluation}
\label{sec:eval}

\subsection{Intermediate Metric}
\label{sec:intermediate_metric}

As discussed in Sec.~\ref{sec:intro}, 
although the multi-stage training paradigm offers many advantages, the optimization objectives of the multimodal model and the downstream CTR model differ. 
Therefore, it is necessary to select a reliable intermediate metric to guide the optimization of the multimodal model. Based on extensive experimentation, we identify image-based search recall as this critical intermediate metric. Below we explain the rationale and justify this choice.

To evaluate how improvements in an intermediate metric translate into downstream gains, 
we define a metric called \textit{exchange rate}, computed as the ratio between the improvement in the downstream metric and the improvement in the intermediate metric. 
Specifically, in our experiments the exchange rate is the ratio of the change in AUC for the downstream CTR model to the change in a given intermediate metric.
For example, a +1\% increase in an intermediate metric that yields a +0.1\% increase in AUC corresponds to an exchange rate of 1.0. 
By this definition, an effective intermediate metric should ideally have a higher exchange rate, or at least remain consistently positive, because a negative exchange rate would indicate that intermediate improvements do not translate to downstream gains.

\begin{figure}[h]
  \centering
  \includegraphics[width=0.95\linewidth]{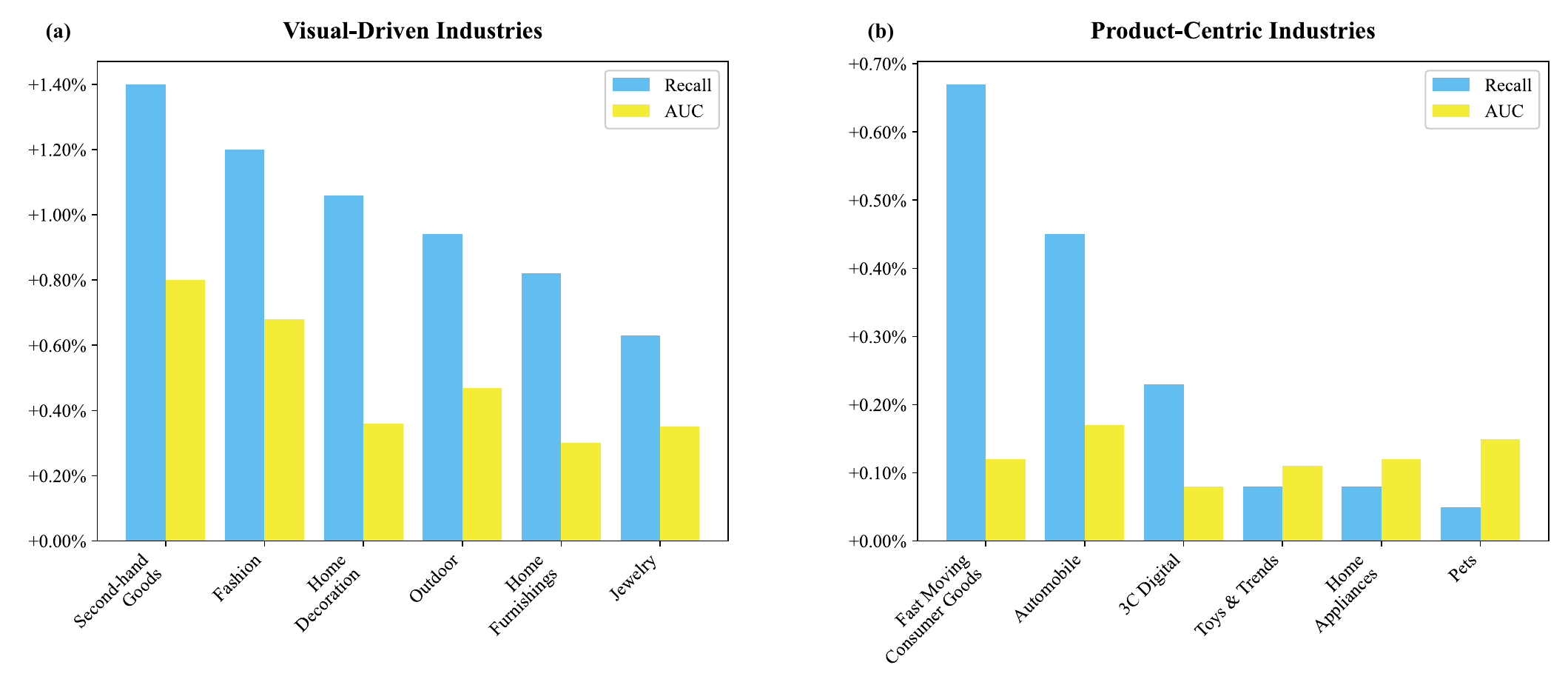}
  \caption{The positive correlation between the image-based search recall and AUC.} 
  \label{fig:recall_auc}
\end{figure}

Based on the criterion above, as shown in Fig.~\ref{fig:recall_auc}, image-based search recall emerges as a robust intermediate indicator, characterized by an exchange rate that remains strictly positive across all industries. 
Specifically, Fig.~\ref{fig:recall_auc}(a) illustrates that in visually-driven industries,  where the visual appeal of products matters, recall and AUC exhibit a strong positive correlation. 
This suggests that better multimodal representations lead to larger downstream performance gains.
As shown in Fig.\ref{fig:recall_auc}(b), in industries that prioritize product function, features, and utility, the exchange rate likewise remains consistently positive, indicating that recall and AUC change in the same direction.
In summary, improvements in image-based search recall on our evaluation dataset consistently coincide with increases in AUC for the downstream CTR model.
Furthermore, image-based search recall directly reflects the ability of multimodal representations to capture semantic information.
Higher image-based search recall suggests that the model can accurately identify and match deep cross-modal associations (e.g., between product images and textual descriptions), thereby enabling the construction of a representation space with richer semantics.



In our decoupled three-stage architecture, image-based search recall is pivotal for aligning the goals of multimodal representation learning with those of downstream tasks, serving as a bridge that ensures consistency and coherence across the optimization process.

\subsection{Evaluation on Intermediate Indicator}
\label{subsec:search_eval}


After identifying image-based search recall as the optimization metric for our multimodal model, 
we first report the model's performance on the corresponding offline retrieval tasks. 
Based on the rationale in Sec.~\ref{sec:intermediate_metric}, 
we expect that improvements in this intermediate metric will translate into downstream benefits, as further confirmed by the online A/B test in Sec.~\ref{subsec:online_ab}.

Our multimodal representation model, \model, is designed to emphasize visual understanding and cross-modal matching, with a focus on improving performance in image-driven scenarios. 
Although we also report text-based search results for reference, our core assessment focuses on retrieval performance in image-based search.
This section provides a detailed description of the evaluation dataset, metrics, and experimental results.

\subsubsection{Evaluation Dataset}

The image-based search test set comprises 5K image queries, averaging 200 associated products per query, for approximately 1 million annotated samples.  
The annotators assess each product's relevance to its query, labeling items as \textit{relevant} or \textit{irrelevant}.
For text-based search, the test set comprises 10 million queries matched against a product database of 20 million items.
As an auxiliary evaluation, the text-based test set is not manually annotated but is derived directly from real-world user interaction logs for large-scale evaluation.
For both searches, the task is to retrieve products that are semantically related to the textual or visual query.
Each product includes both an image and a title, providing multimodal information to support accurate matching.

\subsubsection{Evaluation Metrics}

Given an image or text query, the retrieval task ranks all candidate products by their similarity to the query and returns the top-$k$ results. 
The goal is to maximize the number of relevant items in this ranked list, thereby evaluating the model's effectiveness across different search scopes.
To quantify retrieval performance, we use Recall@$k$, which measures the proportion of relevant products among the retrieved top-$k$ items. 
This metric is particularly suitable for e-commerce, where maximizing coverage of relevant items at early ranks is critical for user engagement and business outcomes.

\subsubsection{Experimental Results}

In our experiments, \model achieves strong performance on both image- and text-based search tasks.
As shown in Tab.~\ref{tab:eval_recall}, for image queries \model achieves 95.1\% recall at Top-1, indicating high accuracy in single-item retrieval. 
Although recall decreases as $k$ increases, it remains above 90\% at Recall@50, demonstrating robust retrieval capability as the return scope expands.
Furthermore, \model maintains stable performance from Recall@1 to Recall@10, with values consistently above 94\%, highlighting its effectiveness at retrieving relevant products within a narrow rank range.
For text-based search, \model achieves Recall@1000 of 47.7\%, indicating its ability to identify relevant products from a much larger candidate pool.
The relatively lower recall for text-based search is expected, primarily due to the inherent ambiguity of natural language and the large scale of the retrieval corpus.
Specifically, textual queries often express less precise intent than visual queries, and the large candidate pool inevitably contains many semantically relevant but unclicked items.


Overall, \model demonstrates strong retrieval performance, accurately identifying relevant products for both visual and textual queries. This helps users find desired items more efficiently, thereby improving overall user satisfaction and engagement in e-commerce scenarios.

\begin{table}[h]
    \caption{Performance on image-based search recall.}
    \label{tab:recall_image_search}
    \centering
    \setlength\tabcolsep{7pt}
    \setstretch{1.2}
    
    \begin{tabular}{cccccc}
    \toprule
    \textbf{Metric}  & Recall@1 & Recall@5 & Recall@10 & Recall@20 & Recall@50 \\ 
    \midrule
    \textbf{Value}   & 95.1\%            & 94.9\%            & 94.2\%             & 93.4\%             & 90.8\%             \\ 
    \bottomrule
    \end{tabular}
    \label{tab:eval_recall}
\end{table}

\subsubsection{Qualitative Results}

\begin{figure}[t]
  \centering
  \includegraphics[width=\linewidth]{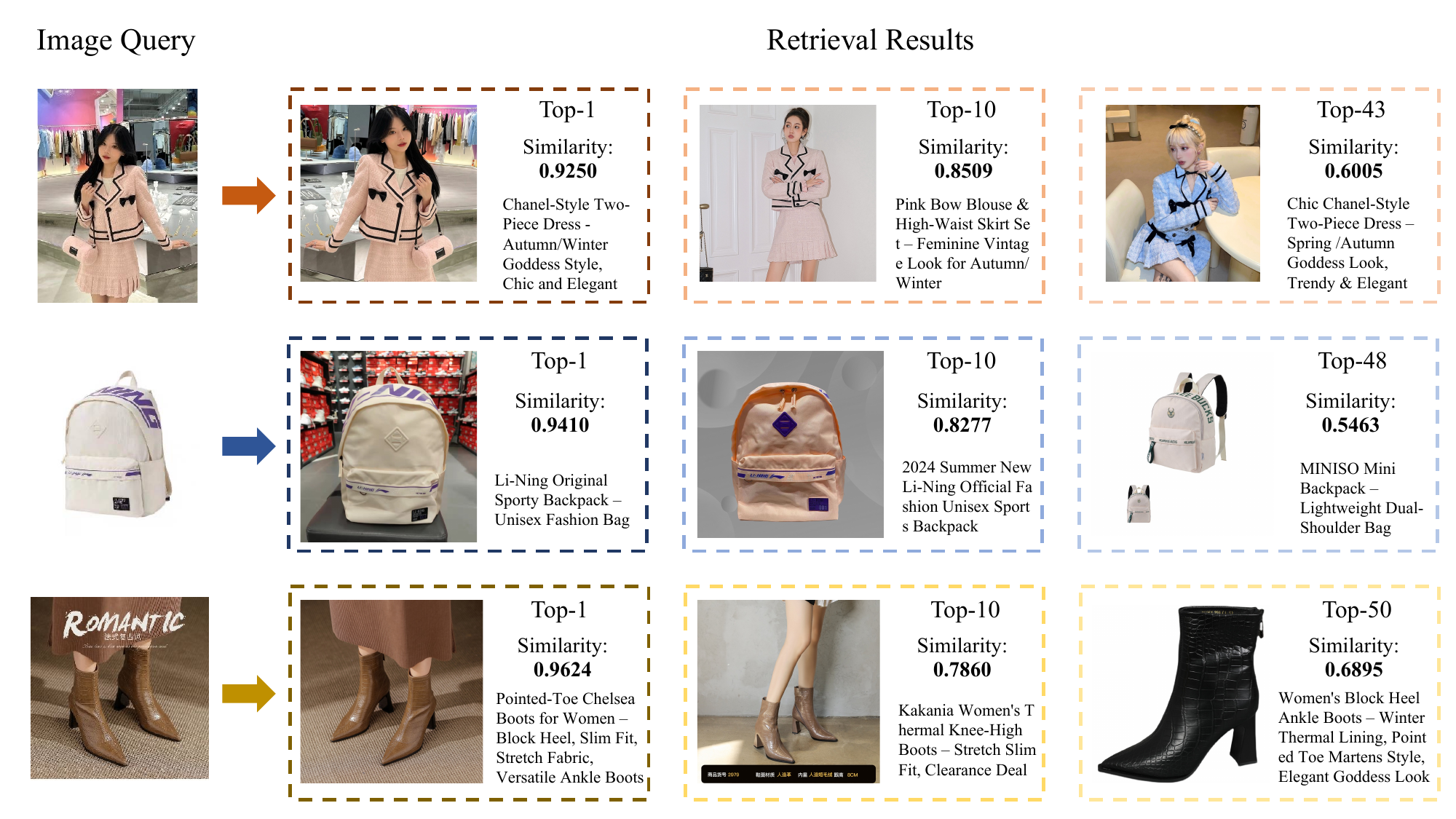}
  \caption{The visualization of image-based search using multimodal representations of \model.} 
  \label{fig:retrieval_vis}
\end{figure}

As illustrated in Fig.~\ref{fig:retrieval_vis}, we conduct a qualitative analysis of \model to further illustrate its strong retrieval capabilities. 
Taking the case in the first row as an example, 
the product characterized by the ``Chanel-Style'' feature occupies the Top-1 position, achieving the highest retrieval similarity score of 0.9250.
The product description emphasizes fashion-design elements, reflecting \model's precise understanding of user intent and demonstrating high accuracy in single-item retrieval.
As the search scope expands, the similarity score at the Top-10 position drops to 0.8509. Nevertheless, \model consistently returns the product aligned with the image query. Furthermore, even at Top-43, where the similarity score falls to 0.6005, the returned products remain semantically consistent with fashion styles, indicating that \model can effectively identify relevant products even within a broader search range.
In summary, these qualitative results show that, given an image query, \model retrieves products whose images and titles consistently reflect the user's underlying intent, delivering highly relevant and semantically aligned search results.

\subsubsection{Cross-modal Alignment Visualization}

\begin{figure}[t]
  \centering
  \includegraphics[width=\linewidth]{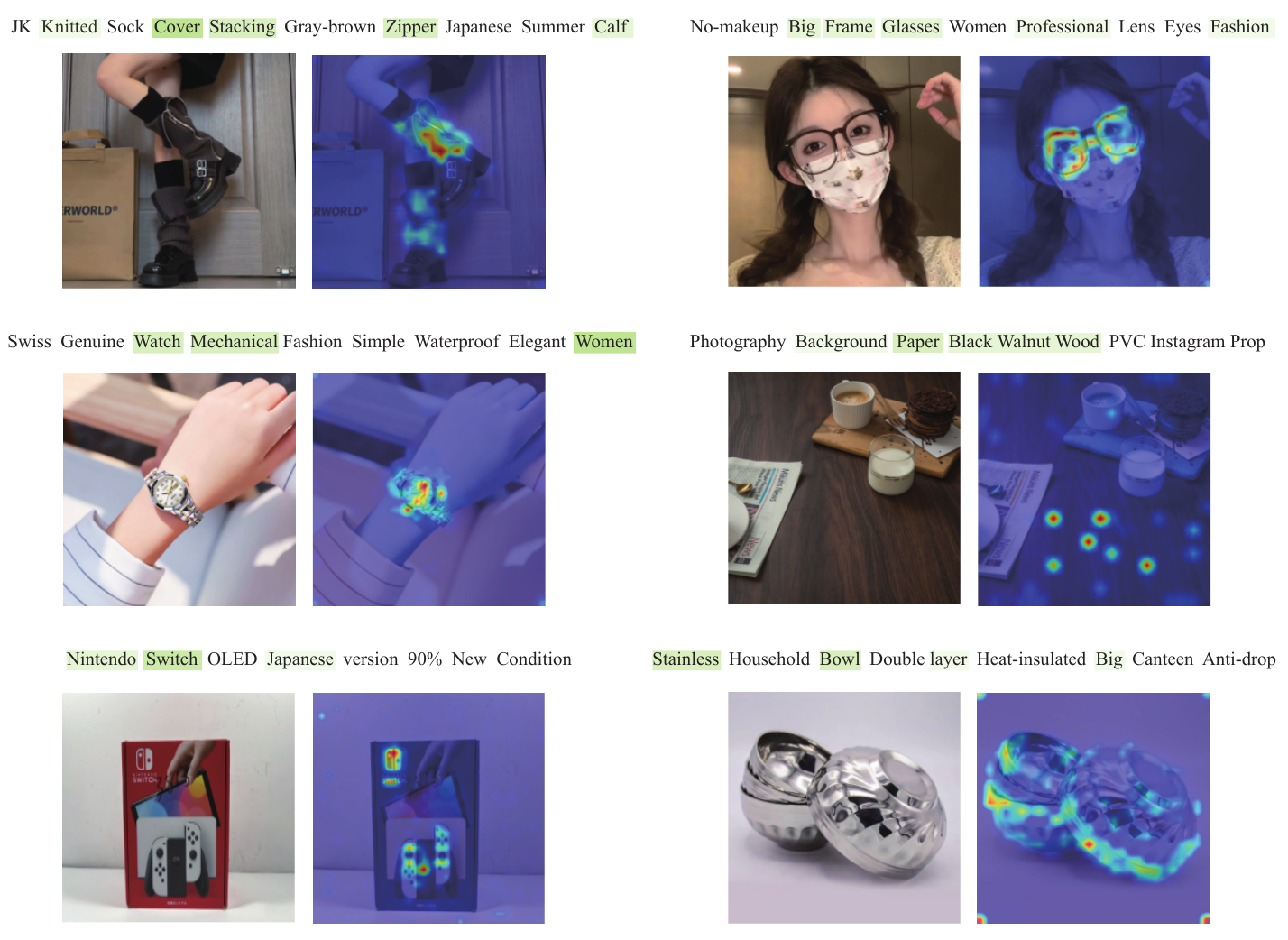}
  \caption{The heatmap visualization of cross-modal alignment of \model.} 
  \label{fig:heatmap}
\end{figure}

To further examine the semantic alignment capability of \model between product images and their corresponding titles, we visualize the regions of attention across both modalities using heatmaps.
As shown in Fig.~\ref{fig:heatmap}, \model successfully highlights the key attributes in both product images and their titles.
The attention maps reveal strong semantic consistency between modalities, indicating that visual and textual information are effectively aligned in a shared representation space.


For example, in the first case the model primarily focuses on the words ``Cover'' and ``Stacking'' in the title, 
while its visual attention concentrates on the sock cover's stacking style in the image.
In the second case, the textual tokens ``Big'' and ``Frame'' align well with the highlighted areas, where the big glasses frame are distinctly emphasized.
Similarly, in the third case, the terms ``Watch'' and ``Mechanical'' in the text corresponds closely to the highlighted mechanical watch in the image. 
The other three examples follow a similar pattern of well-aligned cross-modal attention.
Therefore, this study demonstrates the strong capabilities of \model for cross-modal alignment.

\subsection{Exploration on Scaling Law}
\label{subsec:scaling}

\begin{figure}[h]
  \centering
  \includegraphics[width=0.95\linewidth]{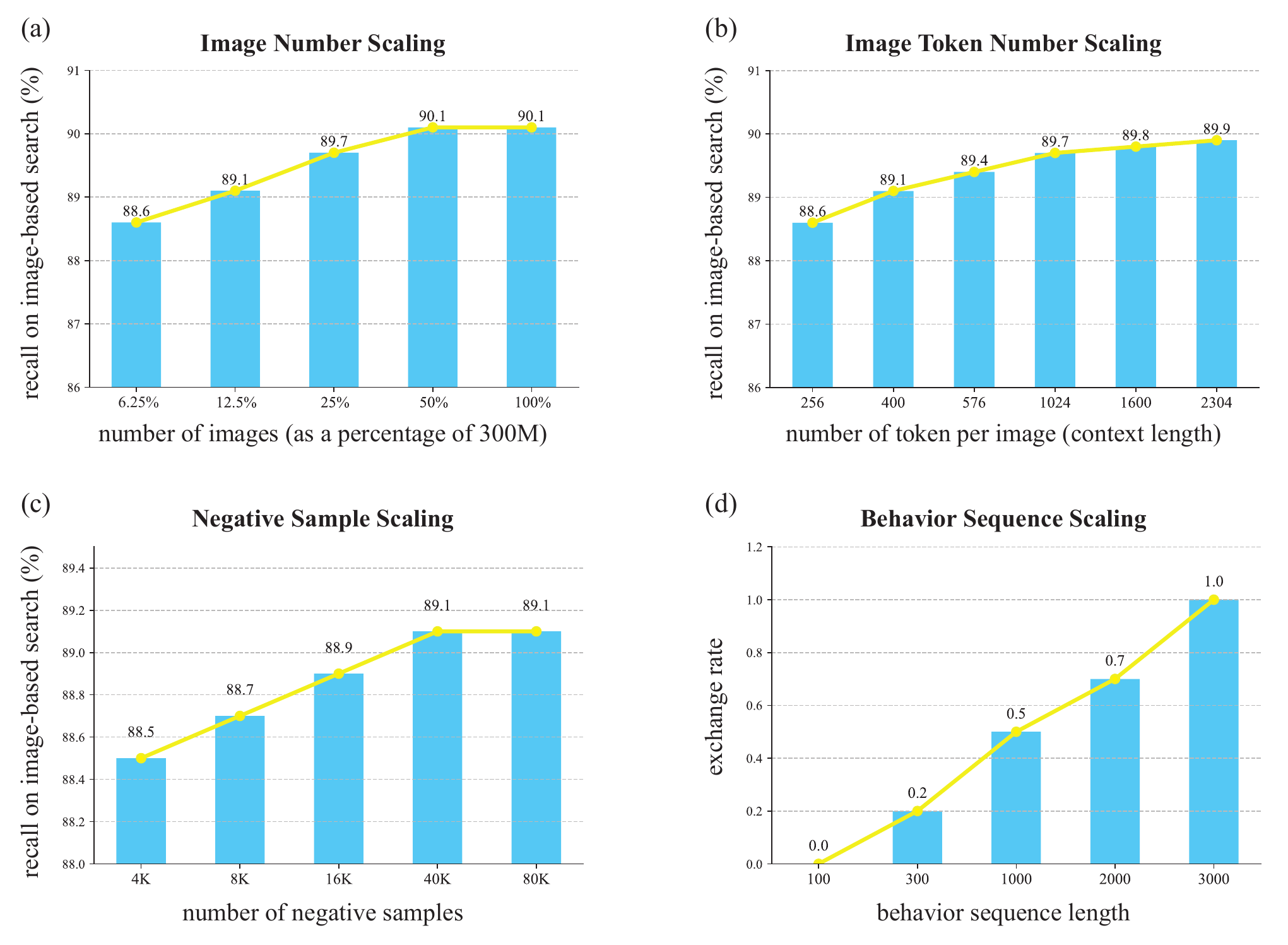}
  \caption{The scaling law in e-commerce search scenarios.} 
  \label{fig:scaling}
\end{figure}

To further understand how multimodal large models affect performance under different scale expansions in e-commerce,
we explore the scaling law from three aspects: the number of training tokens, the number of negative samples, and the length of user behavior sequences.
The image-based search recall is adopted as the evaluation metric. 
As shown in Fig.~\ref{fig:scaling}, the scaling behavior in e-commerce search scenarios follows a clear pattern of diminishing marginal returns.
Given that TBStars-VL is available in both 4B and 14B variants, we adopt TBStars-VL-4B as the base model to balance performance and efficiency.
Therefore, parameter scaling is not discussed in this section.

\subsubsection{Scaling on the Number of Training Tokens}

In the post-training stage, the total number of training tokens equals the number of images in the dataset multiplied by the tokens per image.
Here we investigate how model performance varies with the scale of training tokens.
As shown in Fig.~\ref{fig:scaling}(a), the recall increases as the number of training images grows.
However, marginal gains taper off as dataset size increases, indicating a pattern of diminishing returns. 
Fig.~\ref{fig:scaling}(b) shows a similar pattern when increasing tokens per image: recall improves, but the rate of improvement gradually declines. 
Overall, these results suggest a scaling law linking token count and retrieval performance that exhibits diminishing returns, offering practical guidance for scaling training data in e-commerce search scenarios.

\subsubsection{Scaling on the Number of Negative Samples}

For post-training, we employ a contrastive learning approach with Spatial-Temporal Negative Sampling to enhance the diversity of negative samples. 
Therefore, we also explore how the scale of negative samples affects model performance. 
From Fig.~\ref{fig:scaling}(c), we observe a clear upward trend in recall with the increasing number of negative samples, which can be attributed to the increased diversity of negative samples, enabling the model to learn more comprehensive and discriminative representations. 
However, the recall does not improve when the number of negative samples increases from 40K to 80K, indicating a diminishing marginal effect. 
This suggests that 40K negative samples may already capture sufficient diversity, so further scaling yields little additional information. 
Overall, these findings indicate diminishing marginal returns from negative-sample expansion.

\subsubsection{Scaling on the Length of Behavior Sequence}
\label{sec:eval_behavior_len}

To evaluate how user behavior sequence length affects performance, we use the aforementioned exchange rate as a key metric. 
As illustrated in Fig.~\ref{fig:scaling}(d), as the the length of the behavior sequence increases from 100 to 3000, the exchange rate rises from 0.0 to 1.0. 
The underlying mechanism for this improvement lies in the increased richness of user behavior data. 
That is, longer sequences provide a more comprehensive view of user preferences, enabling the model to learn finer-grained patterns and reduce noise from short-term fluctuations. Additionally, longer sequences allow the model to capture long-term dependencies (e.g., seasonal trends and evolving user interests), which are critical for accurate retrieval and ranking.

This upward trend in the exchange rate reveals the scalability potential of user behavior data. 
The increase in exchange rate with longer sequences suggests that expanding user behavior data can amplify the return on investment from improvements in the intermediate metric. 
Thus, further scaling of behavior sequences could yield additional performance gains, making it a promising direction for future optimization. 
These findings underscore the importance of leveraging extensive user behavior data to maximize retrieval effectiveness in e-commerce.  Moving forward, we will explore even longer sequences, potentially encompassing the lifelong journey of users on Taobao.

\subsection{Online Performance}
\label{subsec:online_ab}

To verify the effectiveness and generalizability of the proposed multimodal representation model \model in e-commerce scenarios, we conduct an evaluation via online A/B testing. 
As mentioned in Sec.~\ref{sec:intro}, in this report, we focus on the CTR as a core metric to measure users' interest in retrieved products. 
The experimental results in 
Tab.~\ref{tab:online_test} show that \model achieves an average CTR improvement of 20.00\%, ndicating a substantial gain in overall search quality and demonstrating \model's effectiveness in enhancing user engagement with recommended content.
To further investigate the \model's advantages in different business scenarios, we present an in-depth analysis from the following three dimensions.

\vpara{New Products.}
The promotion of new products serves as a crucial driver for growth on e-commerce platforms. However, traditional recommendation systems face significant challenges in handling new items due to the lack of historical exposure data. Specifically, ID-based embeddings struggle to construct meaningful feature vectors in the absence of interaction signals, thereby limiting the visibility and matching effectiveness of new products.
In contrast, our proposed content-based multimodal representation does not rely on exposure history. 
Instead, it leverages product content by integrating visual and textual features to produce rich semantic representations. 
This enables a substantial improvement in the discoverability of new products.
The observed 34.80\% uplift in CTR for the new product scenario provides empirical evidence that \model effectively addresses the cold-start problem by enhancing semantic understanding of products.


\vpara{Fashion Category.} 
The fashion products are highly dependent on visual cues and are subject to rapid changes in fashion trends, which impose higher demands on a model's visual understanding ability. \model addresses these challenges by integrating visual and textual information, more accurately capturing users' visual preferences, and thus achieving more precise product recommendations. Experimental results show a 35.74\% increase in CTR for this category, demonstrating \model's distinct advantage on visually dominant products.

\vpara{Bottom-tier Merchants.}
In the e-commerce ecosystem, there exists a common issue where bottom-tier merchants offer highly relevant products but receive limited exposure, primarily due to restricted traffic allocation and less mature advertising experience. 
Consequently, they struggle to reach target customers and need technological support to improve visibility.  
We evaluate \model’s effectiveness for serving such bottom-tier merchants on Taobao. 
As shown in Tab.~\ref{tab:online_test}, the CTR for products from this group increases by 23.03\%, demonstrating that \model can not only identify products with higher relevance in response to user queries, but also enables more effective exposure under tight traffic constraints. 

\begin{table}[h]
    \caption{Online Performance of \model.}
    \label{tab:online_test}
    \centering
    \setlength\tabcolsep{7pt}
    \setstretch{1.2}
    
    \begin{tabular}{cc}
    \toprule
    \textbf{Scenario}          & \textbf{CTR} \\
    \midrule
    Overall    & +20.00\% \\
    \midrule
    New Products      & +34.80\% \\
    Fashion   & +35.74\% \\
    Bottom-tier Merchants & +23.03\% \\
    \bottomrule
    \end{tabular}
\end{table}

The dimension-specific results indicate that \model's excellent performance across business scenarios aligns well with its design objectives. In the new product scenario, \model compensates for ID-based methods' reliance on exposure by leveraging multimodal representations. In the fashion category, it demonstrates a deep understanding of visual semantics, effectively capturing style and trend-related signals. Moreover, for bottom-tier merchants, \model exhibits superior adaptability, esurfacing items that better match user intent even under constrained traffic. In summary, \model shows clear advantages in diverse e-commerce environments, improving the platform's overall performance and exhibiting substantial commercial value. 
\section{Evolutionary Trajectory}\label{sec:history}

\begin{figure}[h]
  \centering
  \includegraphics[width=\linewidth / 100 * 100]{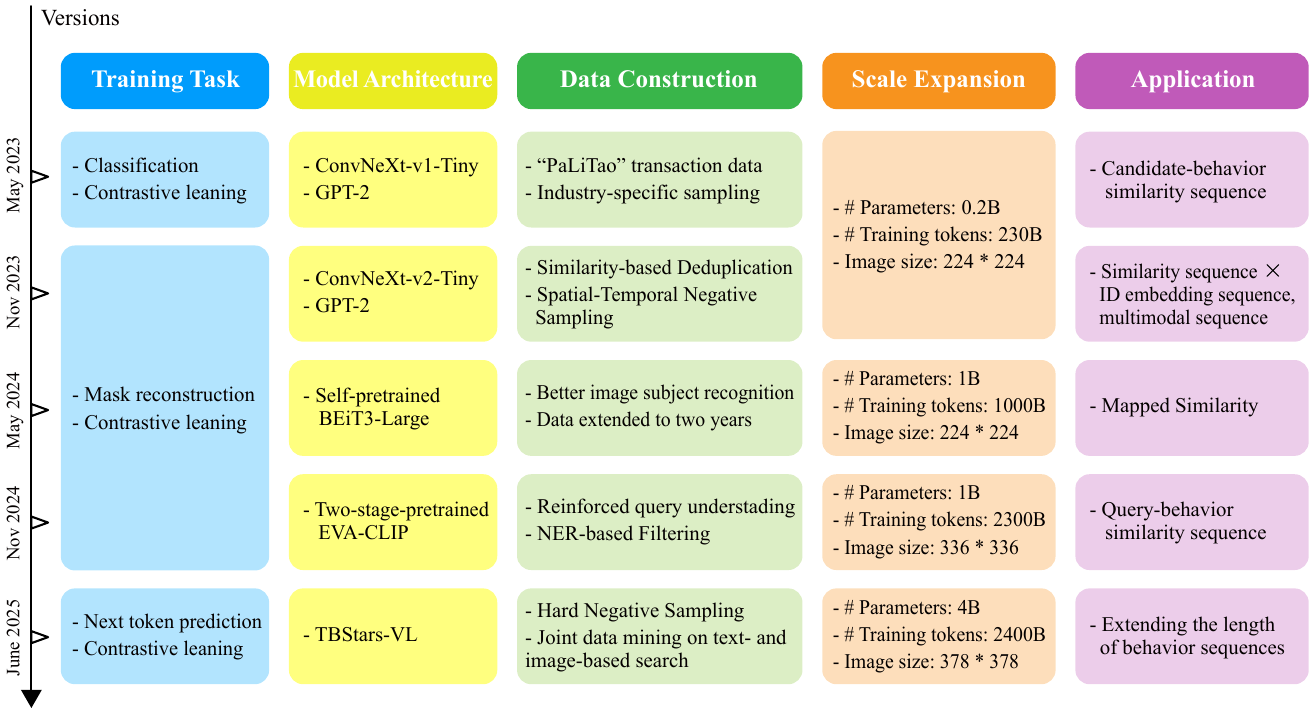}
  \caption{The iteration trajectory of our \model.} 
  \label{fig:history}
\end{figure}

As shown in Fig.~\ref{fig:history}, over the past three years, our \model multimodal representations have undergone five major iterations, 
each characterized by progressive refinements along four aspects:
(1) training task and model architecture; 
(2) data construction and processing strategies; 
(3) scale expansion on model scale, training samples, and image size; and 
(4) downstream applications. 
Together, these five iterations trace a systematic trajectory of innovation that combines increasingly sophisticated learning paradigms, high-quality data pipelines, and scaling laws to enhance multimodal understanding and applicability in e-commerce scenarios. 
In this section, we outline this evolution across these core dimensions.

\subsection{Training Task and Model Architecture }

From a dual-encoder architecture to a generative one, the model design and training objectives of our \model have evolved through continuous iterations, keeping pace with the technical forefront.
\begin{itemize}
    \item 
Version I (May 2023): The initial framework established the three-stage paradigm of ``Pretraining, Post-training, and Application'', 
with pretraining based on classification objectives and post-training driven by contrastive learning. ConvNeXt-v1-Tiny~\citep{liu2022convnet} and GPT-2~\citep{radford2019language} served as the backbone models for visual and textual modalities, respectively.

    \item
Version II (Nov 2023): Inspired by the training of BERT~\citep{devlin2019bert}, the masked reconstruction task was adopted in the pretraining stage, followed by  contrastive learning in the post-training stage. 
We retained ConvNeXt-v2-Tiny and GPT-2 as backbones for their strong ability to understand images and texts at that time.

    \item 
Version III (May 2024): A self-pretrained 1B-parameter BEiT3~\citep{bao2021beit} model was utilized as our multimodal encoder, continuing with masked reconstruction and contrastive tasks to strengthen representation learning.

    \item 
Version IV (Nov 2024): The pretraining stage incorporated a two-stage cross-modal alignment based on EVA-CLIP~\citep{sun2023eva}, and the supervised fine-tuning introduced text-based search samples to reinforce query understanding and improve cross-modal alignment.

    \item 
Version V (June 2025): The novel generative-MLLM-based multimodal representation learning was introduced, and the backbone was upgraded to the state-of-the-art TBStars-VL, 
enabling the deep integration of multimodal understanding with LLMs. 
This latest iteration substantially improved the model’s multimodal understanding capabilities. (Details in Sec.~\ref{sec:method}.)

\end{itemize}

\subsection{Data Construction and Processing}

To ensure the data quality and diversity, the improvements on data processing have been a constant focus across iterations, 
supported by a multi-stage pipeline for data cleaning, mining, sampling, and synthesis.

\begin{itemize}
    \item 
Version I (May 2023): Initially, we utilized the high-quality data samples from ``PaiLiTao'' transactions for supervised fine-tuning, with industry-specific sampling strategies, to ensure robust product modeling and downstream application performance.

    \item 
Version II (Nov 2023): We introduced similarity‑based deduplication and spatial‑temporal negative sampling (see Sec.~\ref{sec:posttrain_2}), which not only significantly improved data quality but also scale the negative sample pool from 4K to over 80K, resulting in enhanced model robustness and more effective representation learning.

    \item 
Version III (May 2024): We optimized the method for image subject recognition and extended the time span of data samples from image-based search to two years, thereby improving data quality, volume, diversity, and temporal coverage.

    \item 
Version IV (Nov 2024): Click‑through data from text‑based searches was leveraged to reinforce query understanding, combined with NER‑based filtering (see Sec.~\ref{sec:posttrain_2}) to reduce textual noise.

    \item 
Version V (June 2025): We apply the hard negative sampling (details in Sec.~\ref{sec:posttrain_2}) on ``PaiLiTao'' transaction data.
We also jointly mine image‑ and text‑based searches to construct higher‑quality multimodal training data with aligned modalities, which further improves discriminative power and applicability across cross‑modal retrieval scenarios.
    
\end{itemize}

\subsection{Scale Expansion}

We have conducted scaling explorations along multiple aspects, including model size, token volume, and image resolution. More experimental results are given in Sec.~\ref{subsec:scaling}.

\begin{itemize}
    \item 
Version I \& II (May-Nov 2023): The baseline model was initially configured with 0.2B parameters, and trained on 230B tokens with image resolution of 224$\times$224.
This setting was maintained in Version II, with efforts focused on refining training objectives and improving data quality.

    \item 
Version III (May 2024): A major scale-up increased model parameters to 1B (5$\times$), training tokens to 1000B (4$\times$), while retaining the image resolution of 224$\times$224.

    \item 
Version IV (Nov 2024): The number of training tokens was further expanded to 2300B, and image resolution increased to 336$\times$336, enhancing the model's ability to capture fine visual details.

    \item 
Version V (June 2025): Parameters surged to 4B (20$\times$), token size reached 2400B, and image resolution was raised to 378$\times$378, 
achieving a deep fusion of multimodality and LLM, significantly enhancing model capacity and its ability to capture fine‑grained content.

\end{itemize}

\subsection{Downstream Application}

Alongside the iterative development of multimodal representations, the methods for applying them to downstream tasks such as CTR prediction have progressed through successive refinements. 
the methods for leveraging them in downstream tasks like CTR prediction have undergone iterative improvements. At their core, these advances enable a more comprehensive modeling of user behavior.

\begin{itemize}
    \item 
Version I (May 2023): To leverage our multimodal representations for CTR modeling, we computed the cosine similarity between the target item and each item in the user behavior sequence.
Then, the semantic similarity sequence was then used as an input feature to the downstream CTR prediction model.

    \item 
Version II (Nov 2023): To strengthen the interaction between multimodal representations and ID embeddings, we enriched the feature set by element-wise multiplication of the similarity sequence with the corresponding item ID embeddings in the user behavior sequence. (Sec.~\ref{sec:application}). 

    \item 
Version III (May 2024): As discussed in Sec.~\ref{sec:application}, 
we applied a learnable linear transformation to the similarity scores, producing values that more clearly distinguished visually similar but non-identical items. This adjustment significantly improved the accuracy of CTR prediction.

    \item 

Version IV (Nov 2024): To support multimodal query understanding, 
we expanded the feature set to include additional query‑based similarity features. 
First, we introduced a sequence of cosine similarities between query and each item in the behavior sequence.
This sequence was used in the same manner as the original similarity sequence, and together they provided richer features to strengthen the downstream CTR prediction. 
Second, a single similarity score between the query and the target item was also included as a new input feature.
In the end, we achieved feature interactions between any two of the four elements: the query, target item, query behavior sequence, and item behavior sequence, further enhancing user preference modeling.

    \item 
Version V (June 2025): To more fully exploit user preference signals in historical behaviors, we expanded the scope of user interests along both temporal and spatial dimensions. In the temporal dimension, we extended the length of user behavior sequences (Sec.~\ref{sec:eval_behavior_len}),
thereby capturing broader temporal patterns of user interest. 
In the spatial dimension, we expanded the user interests by retrieving semantically similar items through nearest-neighbor search.
These expansions capture the temporal dynamics and spatial semantics of user interests, 
laying a foundation for holistic, full-lifecycle user modeling.
%
    
\end{itemize}

In summary, as shown in Fig.~\ref{fig:history}, across these five iterations, the \model multimodal representation model has made significant progress in various areas: training tasks, model architecture, data processing, scale expansion, and downstream applications.
These coordinated advances have not only improved multimodal representation quality but also translated into substantial gains in downstream applications, such as CTR prediction.
The consistent evolution of MOON highlights its strong potential and broad applicability in real-world commercial environments, demonstrating the transformative impact of sustained iteration at the intersection of multimodal representation learning and e-commerce product understanding.

\section{Conclusion}

In this report, we introduce \model, a complete set of sustainable iterative practices for multimodal modeling 
for e-commerce applications.
\model has already been fully deployed across all stages of Taobao search advertising system, 
including retrieval, relevance, ranking, and so on.
The performance gains are particularly significant on click-through rate (CTR) prediction task, which achieves an overall +20.00\% online CTR improvement. 
Through extensive experimentation and five iterative developments, we establish a three-stage paradigm of ``Pretraining, Post-training, and Application'', guided by the image-based search recall metric to align multimodal representation learning with the downstream task. 
Supported by a scalable infrastructure, \model achieves substantial gains in both efficiency and performance.
The latest advancements of our \model have been published in ~\citet{zhang2025moon,nie2025moon2}.
Beyond the CTR prediction, additional applications built on \model are currently under active development for broader e-commerce scenarios.
Looking forward, we believe that the insights and infrastructure established in \model will continue to inspire broader applications across recommendation, search, and advertising systems.


\section{Future Works}

Encouraged by the strong performance of our \model, we believe there are many more promising directions to explore in multimodal representation learning for e-commerce in the future.

\vpara{Data Quality and Scaling.}
We plan to enhance multimodal representation learning by expanding data coverage across more scenarios and modalities, including images, text, video, audio, and ID signals, and by applying large-model-based sample synthesis and feature augmentation. This aims to overcome data quality limitations in scaling and fully realize the benefits of larger models and datasets.

\vpara{Training Paradigm.}
Future work will explore richer multi-stage and multi-task training strategies, integrating mix-of-experts and cross-modal attention architectures to improve fine-grained alignment, model capacity, and generalization while mitigating instability in multimodal joint training.

\vpara{Infrastructure.}
We will advance unified sparse-dense architectures (e.g., RecIS) to accelerate training and inference for 10B+ scale models, improving throughput and efficiency. Downstream sparse models will be optimized for long-sequence (100K+) user modeling and better utilization of multimodal representations.

\vpara{Applications.}
Beyond CTR prediction, multimodal representations will be extended to other stages of recommendation, such as retrieval, relevance, strategic mechanisms, and merchant promotion, with future work detailing these integrations.

\section{Contributors}\label{sec:contributors}

\subsection*{Core Contributors}

\textbf{Algorithm:}
Chenghan Fu, Wanxian Guan, Xiang Zhang, Jianyu Liu, Yueran Liu, Si Chen, Kai Zhang, Daoze Zhang, Yukang Lin, Zhanheng Nie, Jiaqi Liu, Zhiyuan Zeng, Bencheng Yan, Hui Zhao, Pengjie Wang, Jian Xu, Bo Zheng

\textbf{Engineering:}
Hua Zong, Qingtao Zeng, Haolong Guo, Chuyu Huang, Zhendong Li, Shuo Feng, Xiang Li, Jiawei Liu, Qi Wang, Jiawen Liao, Zhengyu Liu, Zexin Yan, Jun Qin, Kaiwen Wei, Fei Wang, Qian Wang, Ju Huang, Siran Yang, Yan Zhang, Huimin Yi, Xiaojie Zhang, Zhenyuan Lai, Yunlong Xu, Qifeng Li, Tongzhen Shao, Jiamang Wang, Peng Sun, Guan Wang, Guang Qiu




\newpage
\bibliography{colm2024_conference}
\bibliographystyle{colm2024_conference}


\end{document}